%% file: EXOT-2017-28-PAPER.tex
\pdfoutput=1
\newcommand*{\ATLASLATEXPATH}{latex/}
\documentclass[cernpreprint, atlasdraft=true, texlive=2016, UKenglish]{\ATLASLATEXPATH atlasdoc}

\usepackage{\ATLASLATEXPATH atlaspackage}

\usepackage{\ATLASLATEXPATH atlasbiblatex}

\usepackage{\ATLASLATEXPATH atlasphysics}

\usepackage{amsmath}
\usepackage{amssymb}
\usepackage[]{multirow}

\graphicspath{{logos/}{figures/}}

\usepackage{EXOT-2017-28-PAPER-defs}

\input{EXOT-2017-28-PAPER-metadata}
\hypersetup{pdftitle={ATLAS document},pdfauthor={The ATLAS Collaboration}}

\begin{document}

\maketitle


\input{Introduction_v8.tex}

\input{Detector_v4.tex}

\input{Models_v6.tex}

\input{DataSamples_v6.tex}

\input{Leptonjets_v7.tex}

\input{EventSelection_v4.tex}

\input{Bkg_v5.tex}

\input{Systematics_v6.tex}

\input{Results_v7.tex}

\input{Conclusions_v4.tex}

\section*{Acknowledgements}
\input{Acknowledgements.tex}

\printbibliography

\clearpage 
\input{atlas_authlist}

\end{document}

%% file: EXOT-2017-28-PAPER-metadata.tex
\AtlasTitle{Search for light long-lived neutral particles produced in \textit{pp} collisions at $\sqrt{s} = 13~\mathrm{TeV}$ and  decaying into collimated leptons or light hadrons with the ATLAS detector}

\AtlasRefCode{EXOT-2017-28}

\AtlasNote{EXOT-2017-28}

\AtlasJournalRef{Eur. Phys. J. C (2020) 80:450}
\AtlasDOI{10.1140/epjc/s10052-020-7997-4}

\AtlasAbstract{%
Several models of physics beyond the Standard Model predict the existence of dark photons, light neutral particles decaying into 
collimated leptons or light hadrons. This paper presents a search for long-lived dark photons produced from the decay of a Higgs boson or a heavy scalar boson and decaying into displaced collimated Standard Model fermions. The search uses data corresponding to an integrated luminosity of \intlumi collected in proton--proton collisions at $\sqrt{s} = 13$~\TeV\ recorded in 2015--2016 with the ATLAS detector at the Large Hadron Collider. The observed number of events is consistent with the expected background, and limits on the production cross section times branching fraction as a function of the proper decay length of the dark photon are reported. A cross section times branching fraction above 4~pb is excluded for a Higgs boson decaying into two dark photons for dark-photon decay lengths between 1.5~mm and 307~mm.
}

\PreprintIdNumber{CERN-EP-2019-140}

%% file: Introduction_v8.tex
\section{Introduction}
\label{sec:intro}

Several extensions of the Standard Model (SM) predict the existence of a dark sector weakly coupled to the SM~\cite{ArkaniHamed:2008qp,Baumgart:2009tn,Falkowski:2014ffa,Curtin:2014cca}.
Depending on the structure of the dark sector and its coupling to the SM, some unstable dark states may be produced at colliders, and could decay into SM particles with sizeable branching fractions.
In order to avoid a new long-range force, a dark Higgs boson is introduced in such scenarios, to give mass to the dark gauge bosons.
The dark Higgs boson may also lead to an exotic decay mode of the Higgs boson, via mixing between the two Higgs sectors, which is one of the favoured production modes that may be probed at the Large Hadron Collider (LHC). This is the mode explored in this search.
Branching fractions of up to 10\% are currently not excluded for Higgs-boson decays into exotic final states~\cite{Aaboud:2019rtt,Sirunyan:2018owy}.
This paper investigates the case where the two sectors couple via a vector portal, in which a dark photon (\gammad) mixes kinetically with the SM photon and decays into SM leptons and light quarks~\cite{Cheung:2009su,Falkowski:2010cm,Falkowski:2010gv}. The kinetic mixing term ($\epsilon$), which can vary over a wide range of values, $\epsilon \sim 10^{-11}$--$10^{-2}$, determines the lifetime of the dark photon. 
For a small kinetic mixing value, the \gammad has a long lifetime, so that it decays at a macroscopic distance from its production point. 
This analysis focuses on small values of the kinetic mixing term, $\epsilon <~10^{-5}$, and a dark photon mass range between twice the muon mass and twice the tau mass.
Due to their small mass, the dark photons are expected to be produced with large boosts, resulting in collimated groups of leptons and light hadrons in a jet-like structure, referred to hereafter as dark-photon jets (\LJs).

The search for displaced \LJs presented in this paper uses the dataset collected by the ATLAS detector during 2015--2016 in proton--proton ($pp$) collisions at a centre-of-mass energy $\sqrt{s}\ =\ 13$~\mytev, corresponding to an integrated luminosity of \intlumi.
The analysis exploits multivariate techniques for the suppression of the main multi-jet background, optimised for the different \LJ channels. This technique allows the exploitation of the fully hadronic signature for the first time in ATLAS \LJ searches, resulting in increased sensitivity compared with previous ATLAS results using the data collected in 2011 and 2012 at 7~and~8~\mytev respectively~\cite{EXOT-2011-12, EXOT-2013-22}.
The results are complementary to those from related ATLAS searches for prompt \LJs using 7 and 8~\mytev data \cite{EXOT-2011-01,EXOT-2012-16,EXOT-2014-09}, which probed higher values of $\epsilon$, and for displaced dimuon vertices using 13~\mytev data \cite{EXOT-2017-03}, which probed higher dark photon mass values. 
Related searches for dark photons were conducted by the CDF and D0 collaborations at the Tevatron~\cite{Aaltonen:2012pu,Abazov:2009hn,Abazov:2010uc} and by the CMS~\cite{CMS-EXO-11-013,CMS-EXO-12-012,CMS-HIG-13-010,Sirunyan:2019xst} and LHCb~\cite{Aaij:2015tna,Aaij:2017rft} collaborations at the LHC. 
Additional constraints on scenarios with dark photons are extracted from, e.g.,  beam-dump and fixed-target experiments~\cite{Blumlein:2011mv,Bjorken:2009mm,Bross:1989mp,Merkel:2011ze,Adlarson:2013eza,Abrahamyan:2011gv,Reece:2009un,Blumlein:2013cua,Gninenko:2012eq,Essig:2010gu,Agakishiev:2013fwl}, $e^{+}e^{-}$ colliders~\cite{Archilli:2011zc,Babusci:2012cr,Aubert:2009cp,Lees:2014xha,Lees:2015rxq,TheBelle:2015mwa,Won:2016pjz,Ablikim:2018bhf,Ablikim:2017aab}, electron and muon anomalous magnetic moment measurements~\cite{Pospelov:2008zw,Davoudiasl:2012ig,Endo:2012hp} and astrophysical observations~\cite{Dent:2012mx,Dreiner:2013mua}.
Given the various constraints, a displaced dark photon with a kinetic mixing term $\epsilon <~10^{-5}$ is allowed for \gammad masses greater than 100~\mymev.

%% file: Detector_v4.tex
\section{The ATLAS detector}
\label{sec:atlas}
ATLAS~\cite{PERF-2007-01} is a multipurpose detector at the LHC, consisting of an inner detector (ID) contained in a superconducting solenoid, which provides a 2~T magnetic field parallel to the beam direction, electromagnetic and hadronic calorimeters (ECAL and HCAL) and a muon spectrometer (MS) that has a system of three large air-core toroid magnets, each composed of eight coils.

 The ID provides measurements of charged-particle momenta in the region of pseudorapidity $|\eta|\leq$~2.5.\footnote{ATLAS uses a right-handed coordinate system with its origin at the nominal interaction point in the centre of the detector and the $z$-axis coinciding with the beam-pipe axis. The $x$-axis points from the interaction point to the centre of the LHC ring, and the $y$-axis points upward. Cylindrical coordinates ($r$,$\phi$) are used in the transverse plane, $\phi$ being the azimuthal angle around the beam pipe. The pseudorapidity is defined in terms of the polar angle $\theta$ as \mbox{$\eta$ = $-$ln tan($\theta$/2)}.} The highest spatial resolution is obtained around the vertex region using semiconductor pixel detectors arranged in four barrel layers~\cite{ATLAS-TDR-2010-19,Abbott:2018ikt} at average radii of 3.3~cm, 5.05~cm, 8.85~cm, and 12.25~cm, and three discs on each side, covering radii between 9~cm and 15~cm.
The pixel detector is surrounded by four layers of silicon microstrips covering radial distances from 29.9~cm to 56.0~cm. These silicon detectors are complemented by a transition radiation tracker (TRT) covering radial distances from 56.3~cm to 106.6~cm. 
%
 
 The ECAL and HCAL calorimeter system covers $|\eta| \leq 4.9$, and has a total depth of 9.7 interaction lengths at \mbox{$\eta$~=~0}, including 22 radiation lengths in the ECAL. The ECAL barrel starts at a radius of 1.41~m and ends at 1.96~m with a $z$ extension of $\pm 3.21$ m, covering the $|\eta| \leq 1.475$ interval. 
In the $1.37 \leq |\eta| \leq 3.2$ region, the ECAL endacap starts at $z \pm 3.70$ m and end at $z \pm 4.25$ m.
The HCAL barrel starts at a radius of 2.28~m and ends at 4.25~m with a $z$ extension of $\pm 4.10$~m, covering the $|\eta| \leq 1.0$ interval. 
In the endcaps regions up to $|\eta| \leq 4.9$, the HCAL starts at $z \pm 4.3$ m and ends at $z \pm 6.05$ m. 

 The MS provides trigger information and momentum measurements for charged particles in the pseudorapidity ranges $|\eta|~\leq~2.4$ and $|\eta|~\leq~2.7$ respectively. It consists of one barrel ($|\eta|~\leq~1.05$) and two endcaps ($1.05~\leq~|\eta|~\leq~2.7$), each with 16 sectors in $\phi$, equipped with fast detectors for triggering and with chambers for reconstructing the tracks of the outgoing muons with high spatial precision.
 The MS detectors are arranged in three stations at increasing distances from the IP: inner, middle and outer. 
Three planes of MS trigger chambers are located in the middle and outer stations. 
The toroidal magnetic field allows precise reconstruction of charged-particle momenta independent of the ID information.

The ATLAS trigger system has two levels~\cite{TRIG-2016-01}, level-1 (L1) and the high-level trigger (HLT). The L1 trigger is a hardware-based system using information from the calorimeters and MS. It defines one or more regions-of-interest (RoI), which are geometric regions of the detector identified by ($\eta$, $\phi$) coordinates, containing interesting physics objects. The L1 trigger reduces the event rate from the LHC crossing frequency of 40~MHz to a design value of 100~kHz.
L1 RoI information provides a seed for the reconstruction of physics objects by the HLT, a software-based system that can access information from all subdetectors. It is implemented in software running on a PC farm that processes the events and reduces the rate of recorded events to 1~kHz.

%% file: Models_v6.tex
\section{Benchmark model}
\label{sec:models}
\begin{figure}[htbp]
\centering
\includegraphics[width=.49\textwidth]{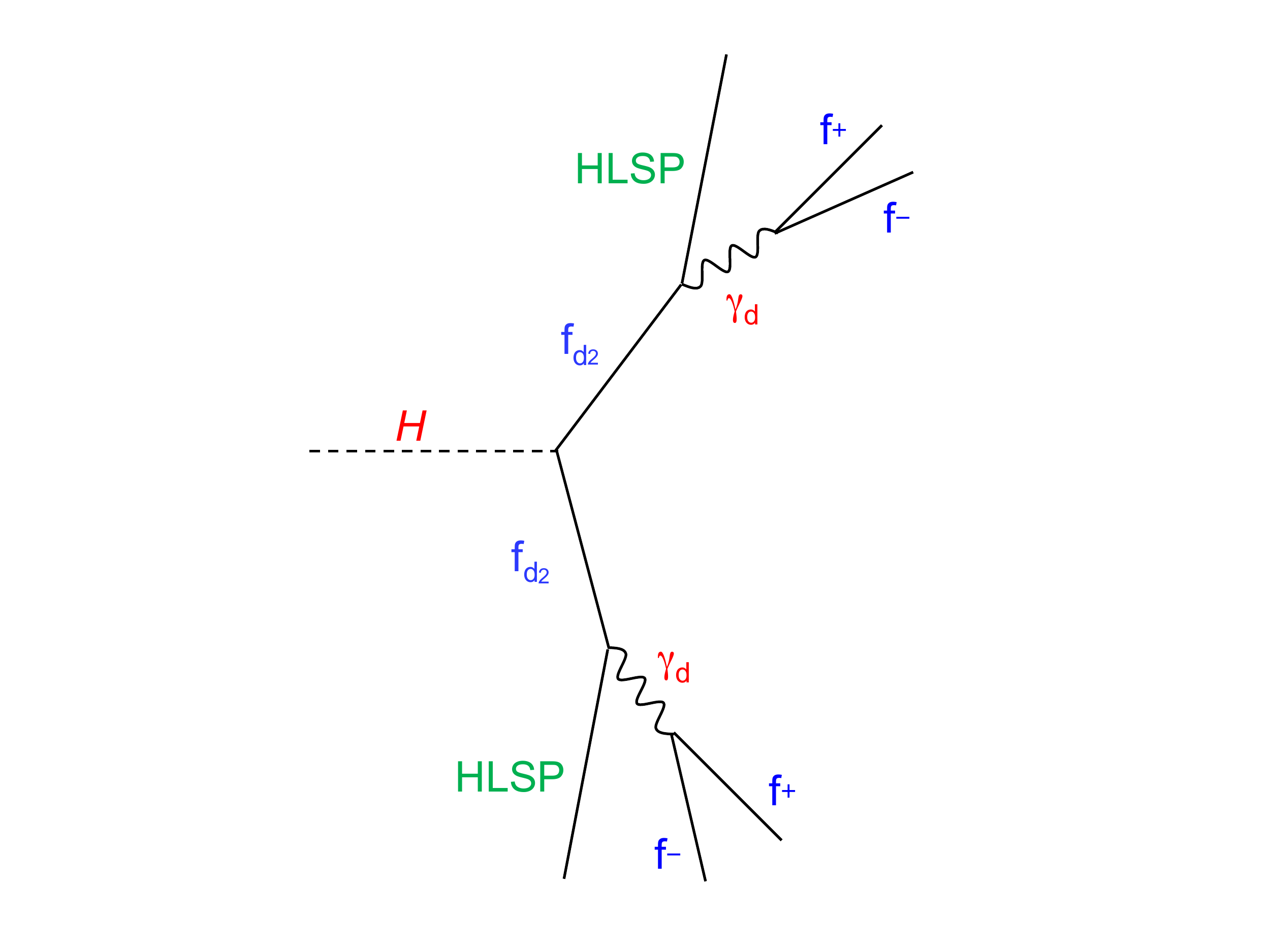}
\includegraphics[width=.49\textwidth]{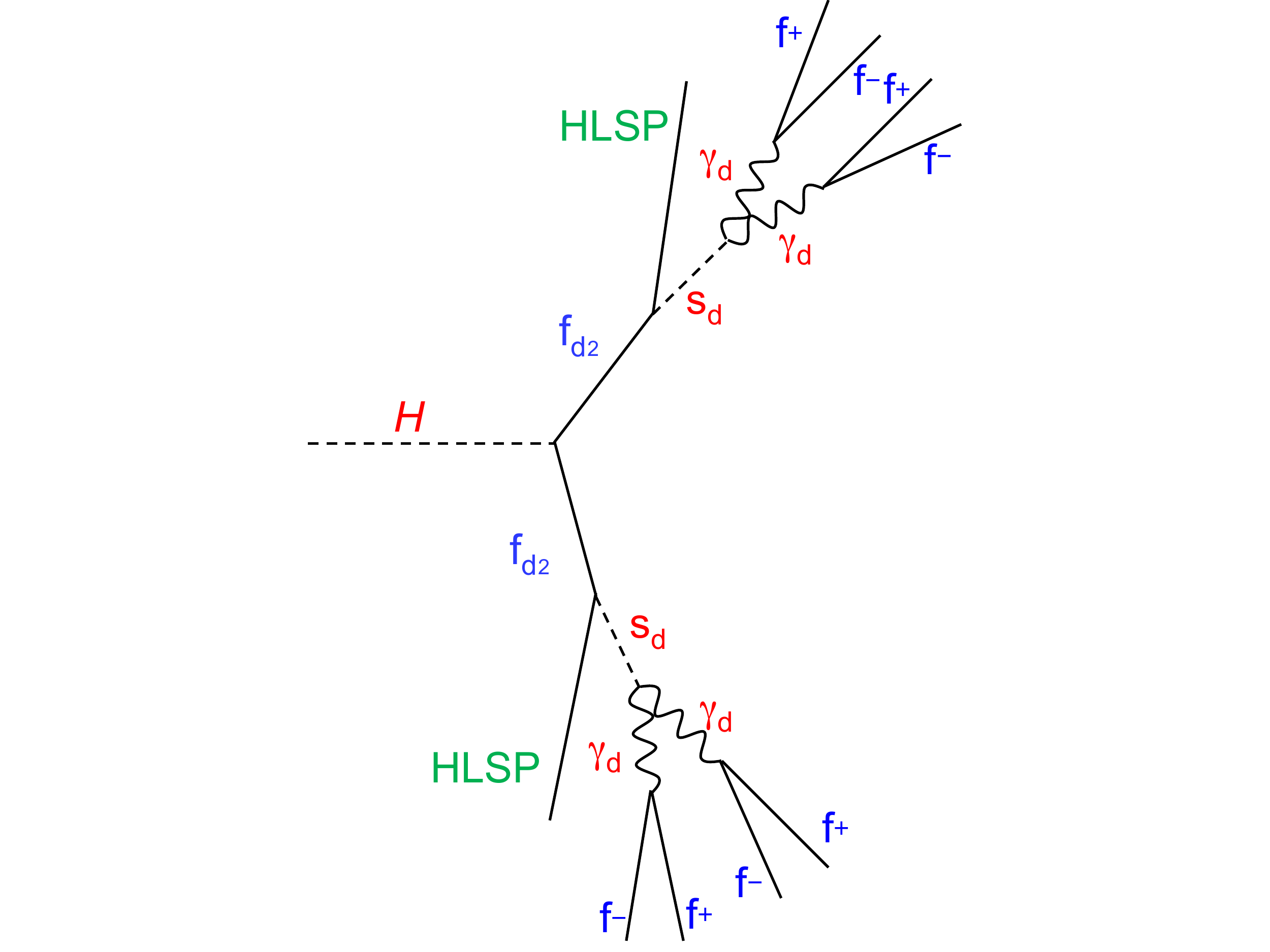}
\caption{The two processes of the FRVZ model used as benchmarks in the analysis. In the first process (left), the dark fermion \fdd decays into a \gammad and an HLSP. In the second process (right), the dark fermion \fdd decays into an HLSP and a dark scalar \sd that in turn decays into a pair of \gammads. The \gammad decays into SM fermions, denoted by $f^{+}$ and $f^{-}$.}
\label{fig:models}
\end{figure}

Among the numerous models predicting dark photons, one class particularly interesting for the LHC features a hidden sector communicating with the SM through the Higgs portal for production and through vector portal for decay. 
The benchmark model used in this analysis is the Falkowski--Ruderman--Volansky--Zupan (FRVZ) model \cite{Falkowski:2010cm, Falkowski:2010gv}, where a pair of dark fermions \fdd is produced via a Higgs boson ($H$) decay.  Two different cases of this model are considered, involving the production of either two or four \gammads. In the first case, shown in Figure \ref{fig:models} (left), each dark fermion  decays into a \gammad and a lighter dark fermion assumed to be the hidden lightest stable particle (HLSP). In the second case, shown in Figure \ref{fig:models} (right), each dark fermion decays into an HLSP and a dark scalar \sd that in turn decays into a pair of dark photons.

In general, dark-sector radiation~\cite{Strassler:2006im} can produce extra dark photons. The number of radiated dark photons is proportional to the size of the dark gauge coupling $\alpha_\textrm{d}$~\cite{Cheung:2009su}.
The dark radiation is not considered in this signal model, which corresponds to an assumed dark coupling $\alpha_\textrm{d} \lesssim 0.01$.
 
The vector portal communication of the hidden sector with the SM is through kinetic mixing of the dark photon and the standard photon 
 
 \begin{equation*}
  \mathcal{L}_{\textrm{gauge mixing}} = \frac{\epsilon}{2}
B_{\mu\nu} b^{\mu\nu},
  \label{eq:lagr}
\end{equation*}

where $B_{\mu\nu}$ and $b_{\mu\nu}$ denote the field strengths of the electromagnetic fields for the SM and dark sector respectively, and $\epsilon$ is the kinetic mixing parameter.
A dark photon with a mass $m_{\gamma_\textrm{d}}$ up to a few \mygev that mixes kinetically with the SM photon will decay into leptons or light mesons, with branching fractions that depend on its mass \cite{Falkowski:2010cm, Meade:2009rb, Batell:2009yf}. 

The mean lifetime $\tau$, expressed in seconds, of the \gammad is related to the kinetic mixing parameter \cite{Cheung:2009qd} by the relation
\begin{equation}
 \tau ~ \propto ~ \Biggl(\frac{10^{-4}}{\epsilon}\Biggr)^2 \Biggl(\frac{100~{\mathrm{MeV}}}{m_{\gamma_\textrm{d}}}\Biggr) ~.
  \label{eq:mixing}
\end{equation}
Equation~(\ref{eq:mixing}) is an approximate expression based on the full relation in Ref.~\cite{Batell:2009yf}.

%% file: DataSamples_v6.tex
\section{Data and simulation samples}
\label{sec:dataSamples}

The analysis presented in this paper uses $\sqrt{s} = 13$~\TeV\ $pp$ collision data recorded by the ATLAS detector during the 2015--2016 data-taking periods. Only runs in which all the ATLAS subdetectors were operating normally are selected. The total integrated luminosities are 3.2~fb$^{-1}$ and 32.9~fb$^{-1}$ for 2015 and 2016 respectively.

Data were collected using a set of dedicated triggers that were active during collision bunch crossings as well as during empty and unpaired bunch-crossing slots.
The LHC configuration for $pp$ collisions contains 3564 bunch-crossing slots per revolution. An empty bunch-crossing is defined as a slot in which neither beam is filled with protons, and in addition is separated from filled bunches by at least five unfilled bunches on each side. Data collected during empty bunch crossings, referred to as the cosmic dataset, are used for the estimation of the cosmic-ray background. The ratio of the number of filled to empty bunch crossings, $F_\textrm{CR}$~=~2.1, is used to scale the number of events in the cosmic dataset to that in the $pp$ collision data. In unpaired bunch crossings, protons are present in only one of the two beams. Data taken during unpaired bunch crossings are used to study characteristic features of beam-induced backgrounds~\cite{DAPR-2012-01} (BIB) and are referred to as the BIB dataset.

Monte Carlo (MC) simulation samples were produced for the model considered in this paper and are summarised in Table~\ref{tab:ggh_FRVZ_param}. 

Samples were generated for the Higgs boson mass of 125~\GeV, and for a hypothetical beyond-the-SM (BSM) heavy scalar boson with a mass of 800~\GeV, considering only the dominant gluon--gluon fusion (ggF) production mechanism. 
The ggF Higgs boson production cross section in $pp$ collisions at \myrts = 13~\tev, estimated at next-to-next-to-leading order (NNLO)~\cite{Dittmaier:2011ti,Harlander:2002wh,Anastasiou:2002yz,Ravindran:2003um}, is $\sigma_{\textrm{\SMs}} = $ 43.87~pb for $m_{H}=$~125~\mygev.
The BSM heavy scalar with a mass of 800~\gev~production cross section is conventionally assumed to be $\sigma = 5~\textrm{pb}$.

The mass of the hidden fermion $m_{f_{\mathrm{d_2}}}$ and of the hidden scalar $m_{s_{\mathrm{d}}}$  were chosen to be low relative to the Higgs boson mass.
Due to the production from a two-body decay of the Higgs boson generated at rest in the transverse plane, events with two back-to-back \LJs are expected. This is also the case leading to four \gammads where each \LJ consist of two collimated dark photons.


The dark-photon mass was chosen to be 0.4~\mygev, above the pion pair mass threshold, and the \gammad decay branching fractions (\BR) are expected to be  \BR(\gdee)~=~45\%, \BR(\gdmumu)~=~45\%, \BR(\gdpaipai)~=~10\%~\cite{Falkowski:2010cm}.
In the generated samples, the proper decay length \ctau of the \gammad was chosen such that  $\sim$80\% of the decays occur in the volume delimited by the muon trigger chambers (i.e.\ up to 7~m in radius and 13~m along the $z$-axis).  Since the analysis is sensitive to a wide range of mean proper lifetimes, a weighting method is used to extrapolate the signal efficiency to other mean proper lifetimes.

All MC samples described above were generated at leading order using \MADGRAPH5\_\AMCatNLO2.2.3
~\cite{Alwall:2014hca} interfaced to \mbox{\textsc{Pythia 8.210}}~\cite{Sjostrand:2014zea} for parton shower generation. The A14 set of tuned parameters (tune) for parton showering and hadronisation~\cite{ATL-PHYS-PUB-2014-021} was used together with the NNPDF2.3LO parton distribution function (PDF) set~\cite{Ball:2012cx}.

\renewcommand{\arraystretch}{1.2}
\begin{table}[ht!]
\caption{Parameters used for the Monte Carlo simulations of the benchmark model.}
\small
\centering
\begin{tabular}{| c | c | c | c | c | c | c | c |}\hline
Sample & $m_{H}$ & $m_{f_{\mathrm{d_2}}}$ & $m_{\mathrm{HLSP}}$ & $m_{s_{\mathrm{d}}}$  & $m_{\gamma_{\mathrm{d}}}$ & $c\tau$ \\
    & $[\gev]$  &$[\gev]$ &$[\gev]$ & $[\gev]$ & $[\gev]$ & [mm]  \\\hline \hline
\gghHtwogd & 125 & 5.0 & 2.0 & -- & 0.4 & 49.23  \\\hline
\gghHfourgd & 125 & 5.0 & 2.0 & 2.0 & 0.4 & 82.40  \\\hline
\gghHtwogd & 800 & 5.0 & 2.0 & -- & 0.4 & 11.76  \\\hline
\gghHfourgd  & 800 & 5.0 & 2.0 & 2.0 & 0.4 & 21.04 \\\hline
\end{tabular}
\label{tab:ggh_FRVZ_param}
\end{table} 

One of the main SM backgrounds in this analysis is multi-jet events. Such events were simulated to perform background studies and to evaluate systematic uncertainties. The MC samples were generated with \mbox{\textsc{Pythia 8.210}} using the same tune and PDF as for the signal samples.

Potential sources of background also include \textit{W}+jets, \textit{Z}+jets, \ttbar, single-top-quark, \textit{WW}, \textit{WZ}, and \textit{ZZ} events. Simulation samples are used to study these backgrounds. The {\textit W}+jets, {\textit Z}+jets , \textit{WW}, \textit{WZ}, and \textit{ZZ} events were generated using \textsc{Sherpa}~2.2.2~\cite{sherpa} with the NNPDF~3.0~NNLO~\cite{Ball:2014uwa} PDF set. Single-top-quark and \ttbar MC samples were generated using \textsc{Powheg-BOX 1.2856}~\cite{Nason:2004rx,Frixione:2007vw,Alioli:2009je,powhegtt} and \mbox{\textsc{Pythia 6.428}}~\cite{Sjostrand:2006za} with the Perugia2012~\cite{perugia} tune for parton showering and hadronisation, and CT10/CTEQ6L1~\cite{ct10,cteq6l1} PDF sets. 

Data and MC samples of $J/\psi \to \mu\mu$ events are used to evaluate systematic uncertainties in muon trigger and reconstruction efficiencies. The MC sample was generated using \textsc{Pythia8+Photos++}~\cite{Golonka:2005pn} with the A14 tune for parton showering and hadronisation, and the CTEQ6L1 PDF set.
The $J/\psi \to \mu\mu$ data sample was selected in 2015--2016 $pp$ collisions using the triggers described in Ref.~\cite{PERF-2015-10}.

The generated MC events were processed through a full simulation of the ATLAS detector geometry and response~\cite{SOFT-2010-01} using the \textsc{Geant4}~\cite{Agostinelli:2002hh} toolkit.
The simulation included multiple $pp$ interactions per bunch crossing (pile-up), as well as the detector response to interactions in bunch crossings before and after the one producing the hard interaction. To model the effect of pile-up, simulated inelastic $pp$ events were overlaid on each generated signal and background event. The multiple interactions were simulated with \textsc{Pythia 8.210} using the A2 tune~\cite{ATL-PHYS-PUB-2012-003} and the MSTW2008LO PDF set~\cite{Martin:2009iq}.

%% file: Leptonjets_v7.tex
\section{Definition of the dark-photon jets}
\label{sec:lj_sele_reco}

\subsection{Dark-photon jet classification}
\label{sec:LJclus}
Displaced \LJs are reconstructed with criteria that depend on the \gammad decay channel. A \gammad decaying into a muon pair is searched for by looking for two closely spaced muon tracks in the MS, while a \gammad decaying into an electron or pion pair, given the high boost of the \gammad, is searched for as an energy deposit in the calorimeters identified as a single narrow jet. MC simulations show that \LJs containing two \gammads both decaying into an electron or pion pair are reconstructed as a single jet.

Tracks that are reconstructed in the MS and are not matched to any track in the ID are used to identify displaced \gammad decays into muons. Since the ID track reconstruction in ATLAS~\cite{ATL-PHYS-PUB-2015-051} requires at least one hit in one of the two innermost pixel layers,
this analysis is sensitive only to displaced \gammad decays occurring after the first pixel layers. 
The search is limited to the pseudorapidity interval $|\eta| < ~2.5$, corresponding to the ID coverage, to ensure that selected muons are isolated from ID tracks. 
Muons with pseudorapidity in the range $1.0 \leq | \eta | \leq 1.1$ are rejected to avoid the transition region of the MS between barrel and endcap.
In order to reconstruct \gammad decays that occur outside of the innermost layer of muon chambers but before the first MS trigger chamber, muons are required to have at least one hit in two of the three MS tracking station. 

Jets used in this search are reconstructed from clusters~\cite{PERF-2014-07} of energy deposits in the ECAL and HCAL using the anti-$k_{t}$ algorithm~\cite{Cacciari:2008gp,Cacciari:2011ma} with radius parameter $R = 0.4$.  The search is limited to \gammad decays into electron or hadron pairs in the hadronic calorimeter. Jets produced in the HCAL are expected to be isolated from activity in the ID, with a high ratio of energy deposited in the HCAL ($E_\mathrm{HCAL}$) to energy deposited in the ECAL ($E_\mathrm{ECAL}$), and appear narrower than ordinary jets.
The standard jet-cleaning requirements~\cite{ATLAS-CONF-2015-029} applied in most ATLAS analyses reject jets with high values of $E_\mathrm{HCAL}$/$E_\mathrm{ECAL}$. A dedicated cleaning algorithm for jets created in the HCAL is applied instead, with no requirements on the ratio $E_\mathrm{HCAL}$/$E_\mathrm{ECAL}$. Jets are required to have transverse momentum \mypt $\geq$ 20~\mygev and $|\eta|~<~2.5$. 
In addition, the weighted time of the energy deposit in the calorimeter cells is required to be in the range [--4~ns,~4~ns] of the expected arrival time for particles produced at $t~=~0$ (bunch-crossing time) and moving with the speed of light, to reduce cosmic-ray background and BIB jets. 

%
\LJs are classified according to the number of muons and jets found within a given cone of angular size $\Delta R\equiv\sqrt{(\Delta\phi)^2 + (\Delta\eta)^2}$ around a muon or jet candidate with the highest transverse momentum.  The cone size is fixed to \mydelr~=~0.4, since the MC simulations show that this selection retains up to 90\% of the dark-photon decay products in the \gghHfourgd decay channel with $m_H$ = 125 \mygev. The \LJ classification is summarised as follows:
\begin{itemize} 
\item \textbf{muonic-\LJ (\mLJ)} -- to select \LJs with all constituent \gammads decaying into muons, at least two muons are required and no jets are allowed in the cone.
\item \textbf{ hadronic-\LJ (\hLJ)} -- to select \LJs with all constituent \gammads decaying into electron or pion pairs in the HCAL, one jet is required and no muons are allowed to be in the cone. The electromagnetic fraction of the jet energy, defined as the ratio of the energy deposited in the ECAL to the total jet energy ($E_\mathrm{ECAL}$/$E_\mathrm{total}$), is required to be less than 0.4. This helps reduce the overwhelming background due to multi-jet production. This variable is also used later, as described in section \ref{sec:hadLJ}
\end{itemize}

Reconstructed \LJs with both muon and jet constituents are not considered in this analysis.


\subsection{Muonic-\LJ selection}
\label{sec:muonLJ}
%
Muonic-\LJs are reconstructed using a Cambridge--Aachen clustering algorithm~\cite{Dokshitzer_1997} that combines all the muons lying within a cone of fixed size in ($\eta$, $\phi$) space.   The algorithm starts from the highest-\mypt muon, searching for additional muons within the \mydelr~=~0.4 cone around the muon momentum vector. If a second muon is found in the cone, the axis of the cone is rotated to the vector sum of the momenta of the two muons, and the search is repeated until no
additional muons are found in the cone. 

Cosmic-ray muons that cross the detector in time coincidence with a $pp$ interaction constitute the main source of background to the muonic-\LJ. The cosmic dataset is used to study this background. A boosted decision tree (BDT) with gradient boosting, implemented in the TMVA framework~\cite{Hocker:2007ht}, is trained to discriminate signal \LJs from the \LJ candidates that originate from cosmic-ray background.
The BDT 
uses the following track variables, for each muon in the \LJ, to classify a \LJ as being from signal or background:
\begin{itemize} 
\item[-] longitudinal impact parameter $z_0$, defined as the minimum separation in the $z$-coordinate between the muon track and the primary vertex (PV);\footnote{The primary interaction vertex is defined to be the vertex with the largest value of $\Sigma p_{\mathrm{T}}^2$, the sum of the squared transverse momenta of all the tracks originating from the vertex.}
\item[-] arrival times measured by the trigger detectors of the MS;
\item[-] pseudorapidity $\eta$;
\item[-] azimuthal angle $\phi$.
\end{itemize} 
Even if the decay is displaced, signal muons point to the primary vertex because of the high boost of the dark photon, resulting in a narrow $z_0$ distribution peaking around zero. By contrast, cosmic-ray muons have a broad $z_0$ distribution.

Cosmic-ray muons mainly come through the two shafts above the ATLAS detector, resulting in two well-defined peaks in the  $\eta$ and $\phi$  distributions. Each hit in the trigger detector of the MS provides a measurement of the time for the muon track, corrected by the time of flight assuming the $pp$ interaction point as the origin of the muon~\cite{MuonBarrel}.
The difference in time measured by the two layers in the middle station and in the outer station is thus useful for discriminating between cosmic-ray muons and collision muons. 
Since cosmic-ray muons are downward going, their arrival times in the layers in the upper part of the MS ($0~<~\phi~<~\pi$) are different from those of collision muons, which are upward-going in this part of the detector. In the lower part of the MS ($\pi~<~\phi~<~2\pi$), cosmic-ray muons and collision muons travel downwards, making hit timing less useful for separating between them.

The cosmic dataset and the signal MC sample \gghHtwogd with $m_H$~=~125~\mygev are used for the training of the BDT. The gain in signal significance obtained from dedicated BDT training with the other signal MC samples is found to be negligible. Figure~\ref{fig:bdtout} (left) shows the BDT output ($\mu$BDT) for the constituent muons of the  \mLJs: the distribution provides a clear separation between signal and background muons from cosmic rays. The $\mu$BDT output is required to be $\mu$BDT~$>~0.21$; the value is chosen to yield the highest signal significance, $S/\sqrt{S+B}$, where $S$ is the number of signal events and $B$ the number of background events.

\subsection{Hadronic-\LJ selection}
\label{sec:hadLJ}
Signal jets are discriminated from multi-jets using a second classifier also based on a BDT (hBDT). The following variables are used as input to the hBDT: 
\begin{itemize} 
\item[-] jet width, defined as the \mypt-weighted sum of the \mydelr between each energy cluster and the jet axis;
\item[-] jet vertex tagger (JVT) output~\cite{ATLAS-CONF-2014-018};
\item[-] $E_\mathrm{ECAL}$/$E_\mathrm{total}$;
\item[-] jet mass, as defined by the jet clustering algorithm~\cite{ATLAS-CONF-2016-035};
\item[-] jet charge, defined as the momentum-weighted charge sum constructed from tracks associated with the jet; tracks are associated with jets using ghost association~\cite{Cacciari:2008gn};
\item[-] jet timing, defined as the energy-weighted average of the timing for each cell in the jet.
\end{itemize}
The JVT is designed to differentiate between pile-up jets and jets originating from the PV. The algorithm uses a multivariate combination of track variables that are sensitive to pile-up. Since jets produced in the hadronic calorimeter have a JVT output distribution similar to that of pile-up jets, the JVT output is used for selection of hadronic-\LJs.
Possible pile-up jets contamination is reduced by the analysis selection to a negligible level.

\begin{figure}[htbp]
\centering
\includegraphics[width=.45\textwidth]{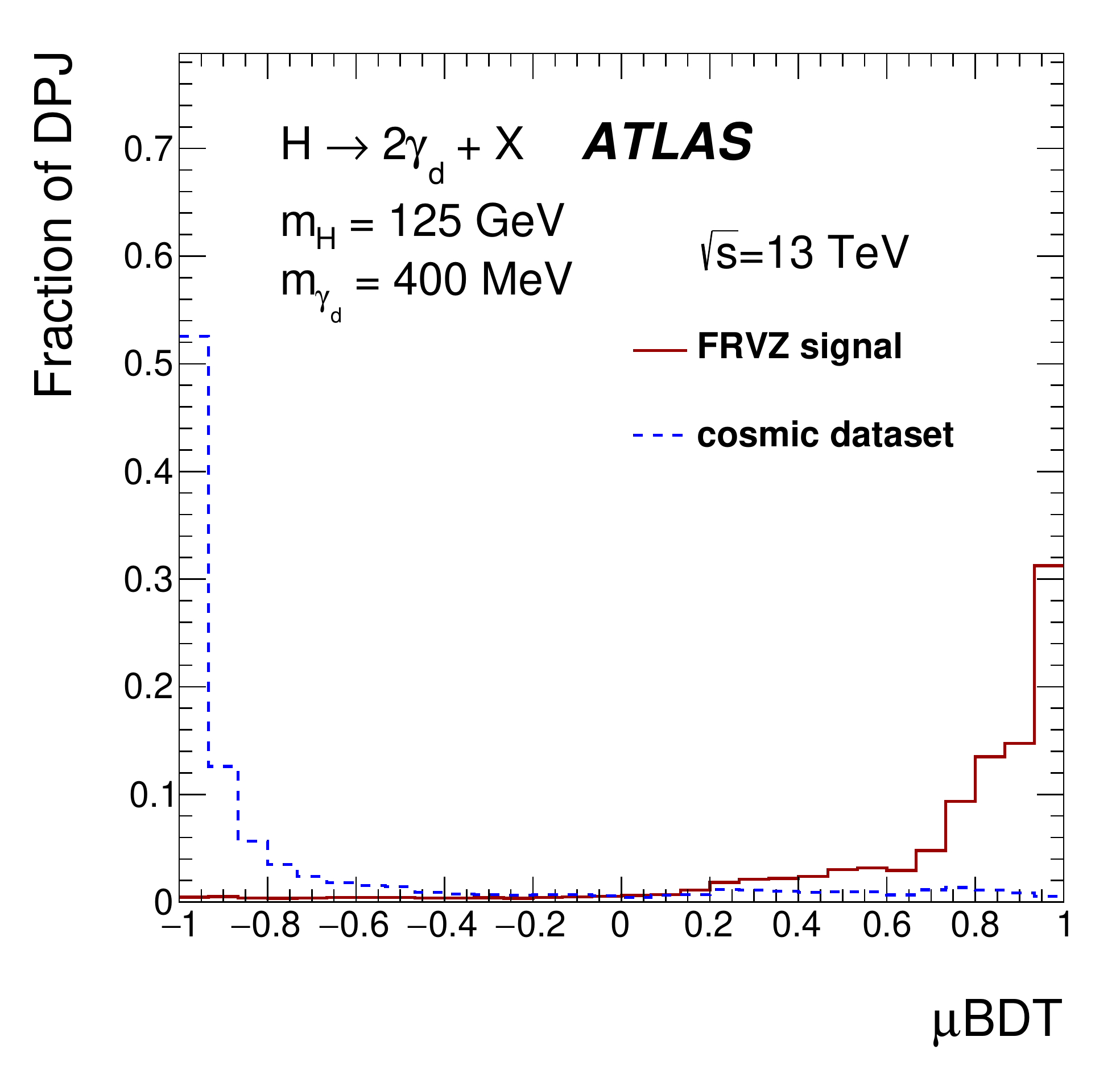}
\includegraphics[width=.45\textwidth]{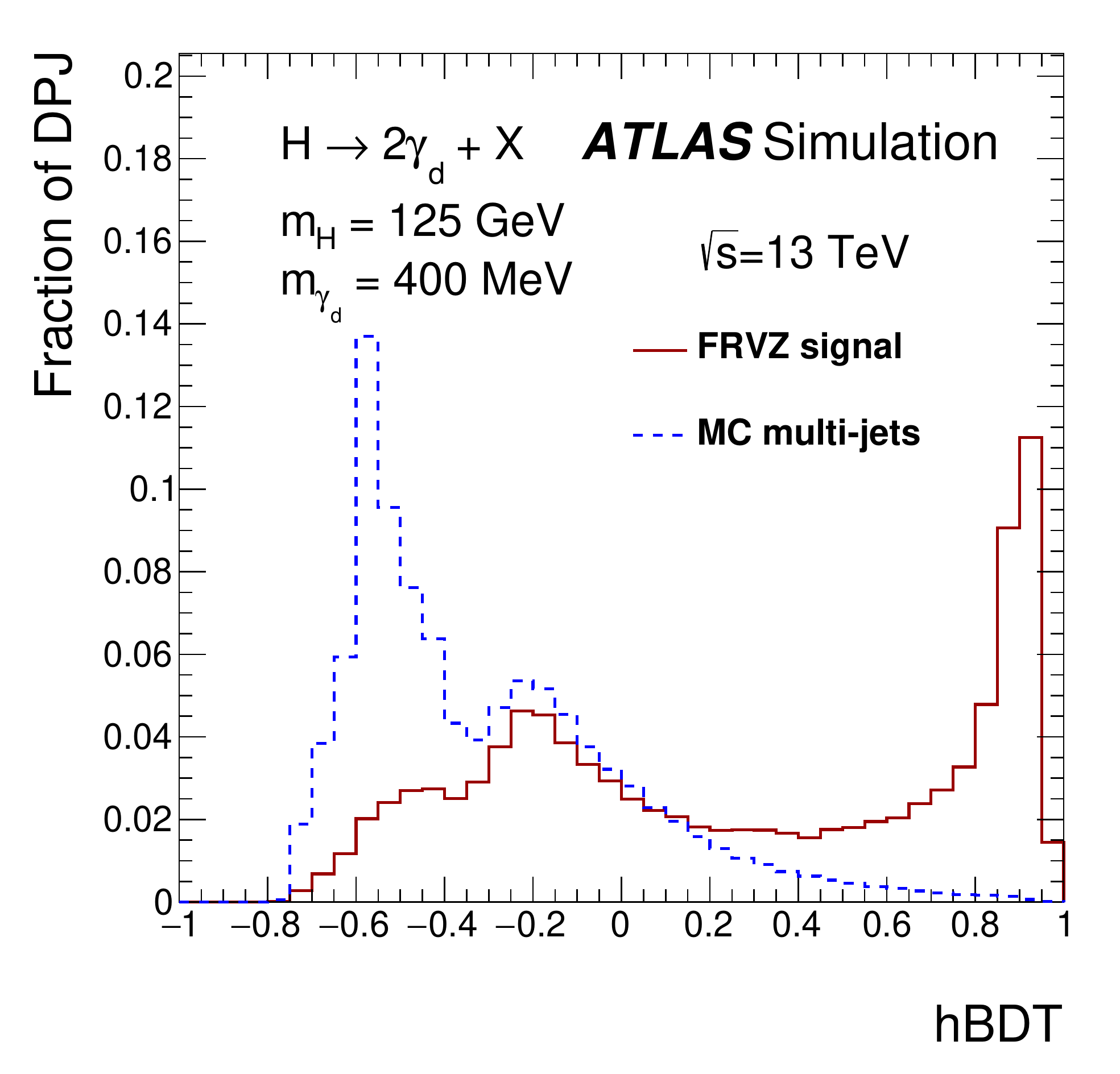}
\caption{BDT output distributions for signal and background for \mLJs (left) and \hLJs~(right). For muonic-\LJs the background is the cosmic dataset and the FRVZ signal sample is the \gghHtwogd process with $m_H$ = 125~\mygev. For hadronic-\LJs the signal MC sample is the \gghHtwogd process with $m_H$ = 125~\mygev and the background is the simulated multi-jet background sample.}
\label{fig:bdtout}
\end{figure}

The signal MC sample \gghHtwogd with $m_H$~=~125~\mygev and the simulated multi-jet background events are used for the BDT training.
The gain in signal significance obtained from dedicated BDT training with the other signal MC samples is found to be negligible.
Figure~\ref{fig:bdtout} (right) shows the BDT output for the \hLJs (hBDT). The peak at ~$\sim$~--0.2 in the BDT  distributions corresponds to jets with a JVT output that indicates a low pile-up probability. The hBDT output is required to be hBDT~$>~0.91$; the value is chosen to yield the highest signal significance. 

%% file: EventSelection_v4.tex
\section{Trigger and event selection}
\label{sec:evtSel}
The standard ATLAS triggers are optimised to select prompt events and are thus usually very inefficient in the selection of displaced objects. 
This search uses events selected by the logical OR of three dedicated triggers targeting displaced objects: two muon triggers and one calorimeter trigger.

The L1 muon trigger used in this analysis requires hits in the middle stations to create a low-$\pT$ $(\geq 6$~\mygev) muon RoI or hits in both the middle and outer stations for a high-$\pT$ $(\geq 20$~\mygev) muon RoI. The muon RoIs have a \mbox{$\Delta\eta\times\Delta\phi$} spatial extent of \mbox{$0.2\times0.2$} in the barrel and of \mbox{$0.1\times0.1$} in the endcaps. L1 RoI information seeds the reconstruction of muon momenta by the HLT, which uses precision-chamber information to confirm or reject the L1 decision.

The first muon trigger, the tri-muon MS-only \cite{TRIG-2012-02},  requires at least three L1 muons with \mypt $\geq$ 6 \mygev in the event, confirmed by the HLT using only MS information. 

The second muon trigger, the muon narrow-scan, is specifically designed to select non-prompt collimated muons originating in the region between the first pixel layer and the first muon trigger plane. It requires an L1 muon with \mypt~$\ge$~20~\mygev confirmed by the HLT using only MS information. At the HLT a `scan' is then performed in a cone of $\Delta R = 0.5$ around this muon, looking for a second muon reconstructed using  only MS information. 
During the course of the 2015--2016 data taking, in order to stay within the allocated trigger-rate limits given the increasing luminosity delivered by the LHC, the \mypt requirement on the second muon was increased from 6~\mygev to 15~\mygev. In addition, both muons were required to be unmatched to any track in the ID, and isolation was required for the leading muon\footnote{The isolation is quantified by summing the \mypt of inner detector tracks with \mypt~>~1~\mygev, excluding the muon candidate, which are found in a cone of $\Delta R$~=~0.2 around the muon candidate.}. This tends to selects events with dark photons of higher \mypt and with more displaced decay position.

The calorimeter trigger, the CalRatio~\cite{TRIG-2012-02}, is designed to select narrow jets produced in the hadronic calorimeter. At L1, the trigger requires a transverse-energy deposit of $E_{\mathrm T}~\geq~60$~GeV within a \mbox{$0.2\times0.2$} (\mbox{$\Delta\eta\times\Delta\phi$}) region in the pseudorapidity range $|\eta|~\leq$~2.4. 
At the HLT, jet reconstruction is then performed with the \antikt\ algorithm using a radius parameter of $R$ = 0.4.  Transverse energy $E_{\mathrm T}~\geq~30$ \mygev and $\log ( E_\mathrm{HCAL} / E_\mathrm{ECAL} )~\geq~1.2$ are required. Jets are required to have no tracks with $p_\mathrm{T}~\geq~2$ \mygev~within \mydelr~$=~0.2$ of the jet axis. Finally, jets are required to pass a BIB removal algorithm that relies on calorimeter cell timing and position. Muons from BIB enter the HCAL and can radiate a bremsstrahlung photon, generating an energy deposit that may be reconstructed as a jet with characteristics similar to the hadronic-\LJ. The algorithm identifies events as containing BIB if the triggering jet has at least four HCAL cells at the same $\phi$ and in the same layer with timing consistent with that of a BIB energy deposit.

Two \LJs satisfying the selection criteria described in Section~\ref{sec:lj_sele_reco} are required in the events selected by the triggers. If more than two \LJs are reconstructed, the one with the highest transverse momentum, labelled the leading \LJ, and the one farthest in \mydelphi\ from the leading one, labelled the subleading \LJ, are used to classify the event. More than two \LJs are found in 9\% of the events in the signal MC sample \gghHtwogd with $m_H$~=~125~\mygev.
Events are classified as one of the three following channels:
\begin{itemize} 
\item {} \mLJ--\mLJ,
\item {} \mLJ--\hLJ,
\item {} \hLJ--\hLJ.
\end{itemize}
In the \mLJ--\hLJ channel, either the \mLJ or the \hLJ may be the leading \LJ. 
%

%% file: Bkg_v5.tex
\section{Multi-jet background estimation}
\label{sec:bkg}
\label{sec:backgroundEstimation}
A data-driven ABCD method is used to estimate the multi-jet background in each of the three channels. The ABCD method uses two nearly uncorrelated variables defined at the event level to create a two-dimensional plane that is split into four parts: region A, where most signal events are located, and three control regions (B, C, and D) that contain mostly background. The number of background events in A can be predicted from the population of the other three regions: $N_{\textrm{A}} = N_{\textrm{B}} \times N_{\textrm{D}}/N_{\textrm{C}}$, assuming negligible leakage of signal into regions B, C and D. For each channel, the ABCD calculation is performed in two separate regions: one background-dominated validation region (VR) to test the validity of the method, and one signal region (SR). 
The SRs are defined by the selection criteria described in Sections~\ref{sec:lj_sele_reco}~and~\ref{sec:evtSel}. These define also the VRs except for the BDT cuts. The VRs BDT cuts for the leading and the subleading \LJs are chosen to have negligible signal contamination, which otherwise can bias the ABCD method validation. SR and VR definitions are summarised in Table~\ref{tab:ABCD_VR_SR}. 

The two event-level variables used to define the ABCD plane are the isolation of the \LJs relative to tracks in the inner detector and the opening angle between the two \LJs in the transverse plane ($|\Delta \phi|$). Displaced \LJs are expected to be highly isolated in the ID. The track isolation ($\sum{p_{\textrm{T}}}$) is defined as the scalar sum of the transverse momenta of the tracks reconstructed in the ID and matched to the PV of the event within a \mydelr = 0.4 cone around the \LJ direction.  Matching to the PV helps reduce the dependence of $\sum{p_{\textrm{T}}}$ on the amount of pile-up. The PV is correctly selected in the $\sim$56\% of the events. However, the selection efficiency does not depend significantly on whether the PV is correctly identified.
The larger of the two $\sum{p_{\textrm{T}}}$ values, \maxsumpt, is used as the event-level variable. For signal, the opening angle $|\Delta \phi|$ is expected to be large, due to production of the \LJs in the two-body decay of a Higgs boson generated at rest in the transverse plane. 

The ABCD method relies on there being only one source of background, or multiple sources that have identical distributions in the ABCD plane.
Muons from BIB originate from beam-halo interactions with the collimators upstream of the ATLAS detector, resulting in muons travelling parallel to the beam-pipe. The analysis requirements select events with two separate energy deposits in the hadronic calorimeter produced by two BIB muons. The requirement $|\Delta \phi|~>~0.1$ in the ABCD plane removes BIB events that  would otherwise contaminate the method for the \hLJ--\hLJ channel, and has no effect on the signal efficiency. After the final selection, the contribution of BIB events to the signal region is negligible.

The ABCD plane is defined for all the three channels by 0~$\le$~\maxsumpt~$\le$~20~\mygev and 0.1~$\le$~$|\Delta \phi|$~$\le~\pi$. The region A is defined by \maxsumpt $< $ 4.5~\mygev and $|\Delta \phi|~>~$ 0.625. Regions B, C, and D are defined by reversing one or both of the requirements: \maxsumpt $> $ 4.5~\mygev and $|\Delta \phi|~>~$ 0.625, \maxsumpt $> $ 4.5~\mygev and $|\Delta \phi|~<~$ 0.625, \maxsumpt $< $ 4.5~\mygev and $|\Delta \phi|~<~$ 0.625 respectively.


\begin{table}[!h]
\centering
\caption{Summary of the definitions of the signal regions (SRs) and validation regions (VRs) used in the ABCD method.}
\begin{tabular}{|l|l|l|}
\hline
Region & Channel & Criteria \\ 
\hline\hline
  \multirow{3}*{SR} & \mLJ--\mLJ & $\mu$BDT > 0.21 for both \LJs \\
                    & \mLJ--\hLJ & $\mu$BDT > 0.21 and hBDT~$>~0.91$ \\
                    & \hLJ--\hLJ & hBDT >~0.91 for both \LJs \\
  \hline
  \multirow{3}*{VR} & \mLJ--\mLJ & --0.75 < $\mu$BDT < 0.35 for leading \mLJ, $\mu$BDT > --0.7 for subleading \mLJ \\
                    & \mLJ--\hLJ & --0.5 < $\mu$BDT < 0.8 and 0.2 < hBDT < 0.8 \\
                    & \hLJ--\hLJ & hBDT <~0.91 for both \LJs\\
\hline
\end{tabular}
\label{tab:ABCD_VR_SR}
\end{table}

In order to estimate the residual cosmic-ray background component in the ABCD plane, the event selection is applied to the cosmic dataset, and the resulting event yield is multiplied by $F_\textrm{CR}$. The expected number of cosmic-ray events in the validation regions is: 4 $\pm$ 3 in region A and 2 $\pm$ 2 in region D for the  \mLJ--\mLJ channel; 10 $\pm$ 5 in region D for the  \mLJ--\hLJ channel. In the signal regions, the expected number of cosmic-ray events is: 8 $\pm$ 4 in region A and 2 $\pm$ 2 in region D for both the \mLJ--\mLJ and the \mLJ--\hLJ channel; 2 $\pm$ 2 in region A for the \hLJ--\hLJ channel. No events are observed in the remaining regions in the cosmic dataset.
The estimated cosmic-ray event yields are subtracted from each of the ABCD regions before using the method to estimate the multi-jet background yield.
Other potential backgrounds to the signal include all the processes that lead to real prompt muons and muons plus jets in the final state, such as the SM production of {\textit W}+jets, {\textit Z}+jets, \ttbar, single-top-quark, \textit{WW}, \textit{WZ}, and \textit{ZZ} events. MC samples are used to study these processes. They give no contribution after the trigger selection and the definition of muonic-\LJ and hadronic-\LJ and do not enter in the ABCD plane.

The signal contamination in the VRs is verified to be less than 5\% for all channels. The linear correlation coefficient between the \maxsumpt\ and $|\Delta \phi|$ variables is verified to be less than 6\% in the VR data, as well as in the SR using multi-jet MC events. The effect of this correlation on the final result is found to be negligible compared to the statistical accuracy.
Table~\ref{tab:abcdInVR} shows the event counts in each of the four regions of the ABCD plane in the validation regions and the expected number of background events in the validation region A in data. Only statistical uncertainties are shown.  The expected contribution from cosmic rays is included in all regions and in the background estimation. The observed number of events in the validation region A is in agreement with the number predicted by the ABCD method within the statistical uncertainties.

As additional validation of the ABCD method, control region D of the SR is divided into four subregions. The subregion with low \maxsumpt and high $|\Delta \phi|$ is treated as a mock signal region, with the other subregions serving as control regions. 
Applying the method, the expected and the observed numbers of events in the mock signal region are: $231\pm58$ and $184$ for the \mLJ--\mLJ channel, $131\pm41$ and $145$ for the \mLJ--\hLJ channel, $402\pm77$ and $479$ for the \hLJ--\hLJ channel.
These are in agreement within the statistical uncertainties.

\begin{table}[h]
\centering
\caption{Event count in each of the four regions of the ABCD plane in the validation regions and expected number of background events in region A. Only statistical uncertainties are shown. The expected contribution from cosmic rays is included in all regions and in the background estimation.}
\small
\begin{tabular}{| c| r | r  | r | r@{$\pm$}l  | r |}
\hline
	\LJ pair type & B & C &  D & \multicolumn{2}{c|}{Expected background in A} & A \\
\hline
\hline
	\mLJ--\mLJ & 4 &  15 & 61     &\phantom{Expected}  20\;&\;10   & 17 \\
\hline
	\mLJ--\hLJ & 455 &  87 & 318  &1611\; &\;227 & 1573 \\
\hline
	\hLJ--\hLJ & 2556 &  536 & 14 &  67\;&\;18  & 57 \\
\hline
\end{tabular}
\label{tab:abcdInVR}
\end{table}
%

\begin{figure}[h]
 \centering
 \includegraphics[width=.45\textwidth]{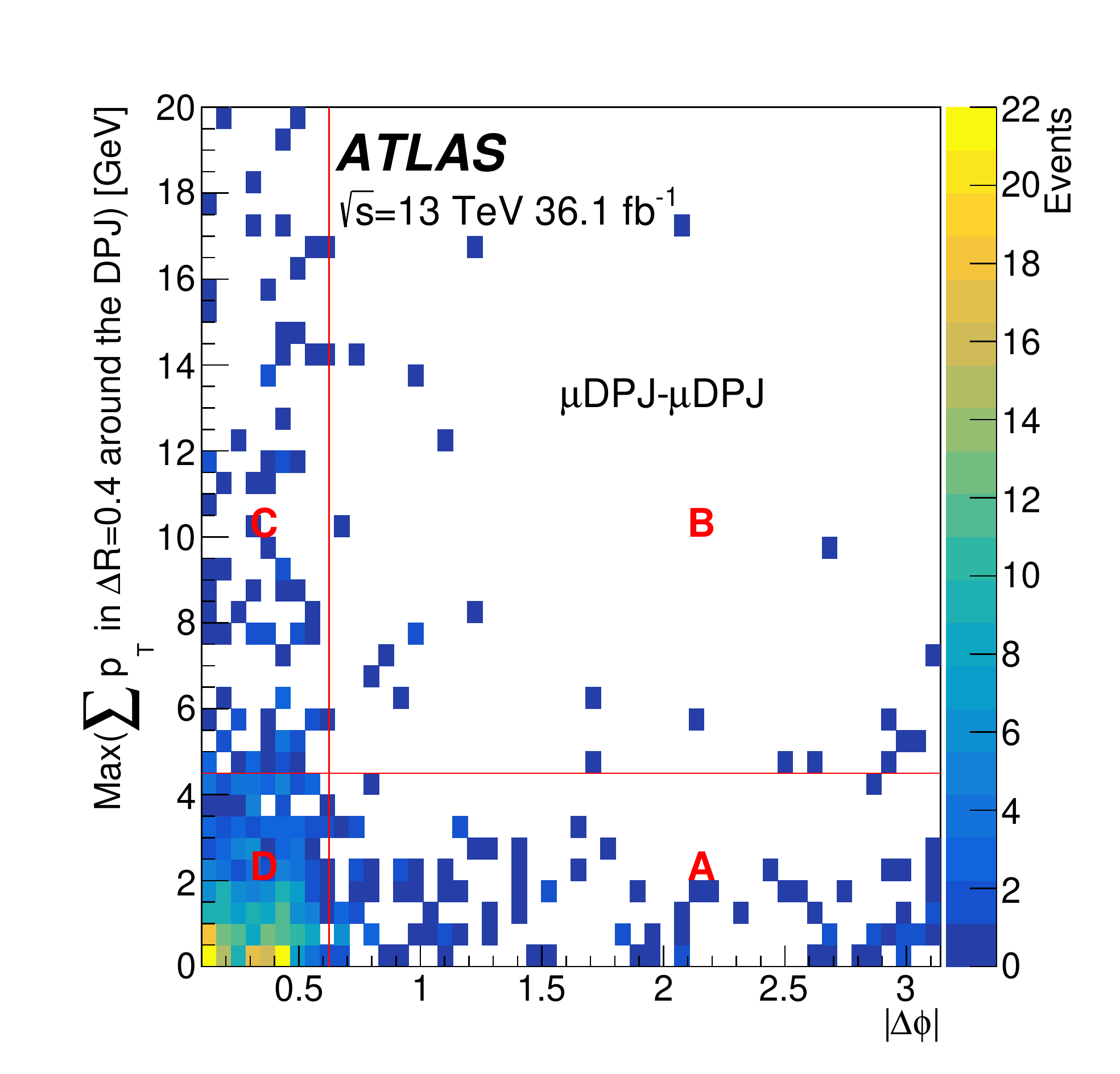}
 \includegraphics[width=.45\textwidth]{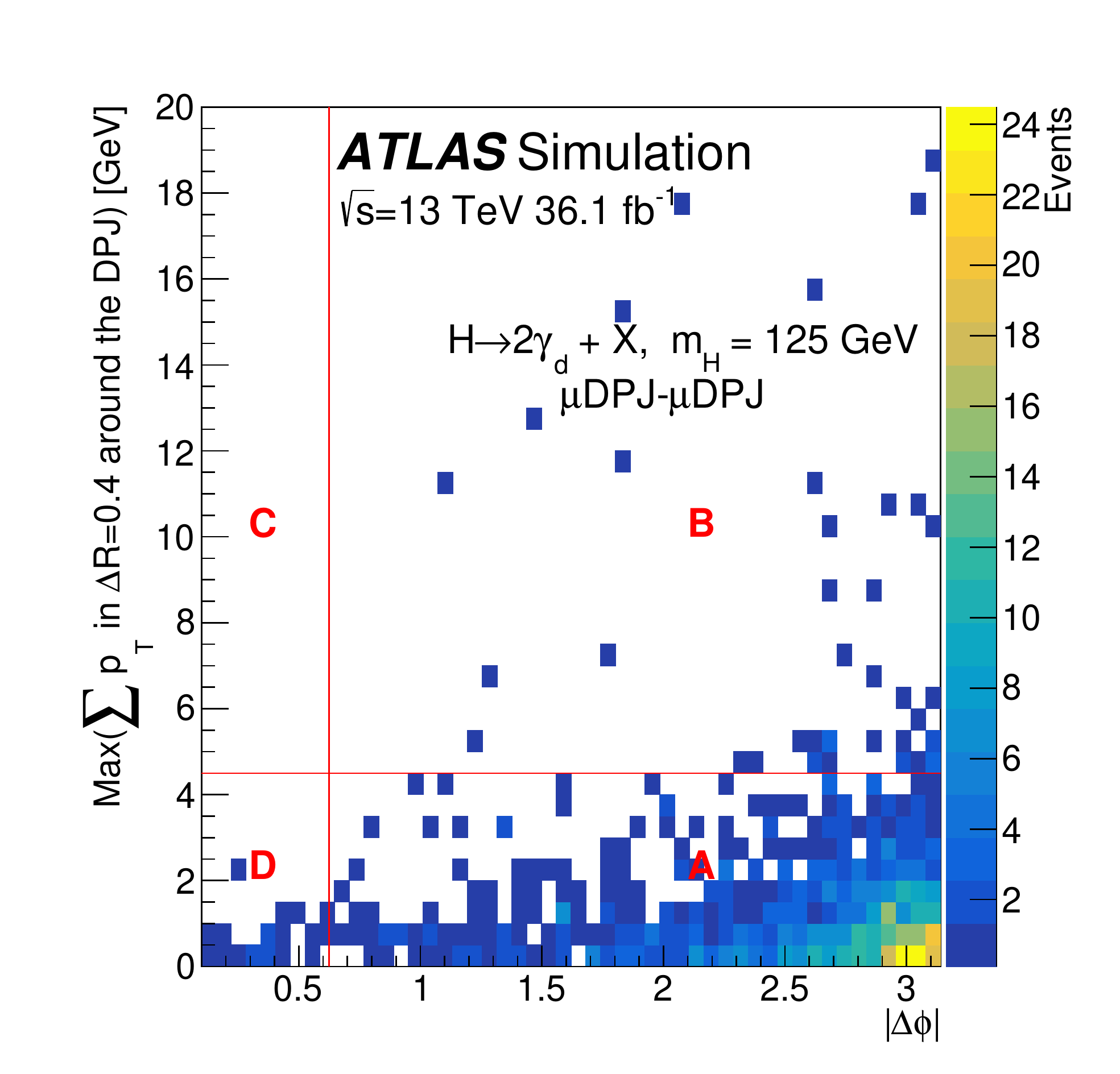}
 \caption{Opening angle between the two \LJs, $|\Delta\phi|$, vs inner-detector isolation, \maxsumpt, in the \mLJ--\mLJ channel for data (left) and MC signal  \gghHtwogd with \mH = 125~\mygev (right), assuming a 10 \% Higgs boson decay branching fraction into \gammad. The red (solid) lines show the boundaries of the ABCD regions.}
 \label{fig:ABCDplane}
 \end{figure}

Figure~\ref{fig:ABCDplane} shows the distribution of events in the ABCD plane of the \mLJ--\mLJ channel in the SR for the collision data and the MC signal  \gghHtwogd with \mH~=~125~\mygev, assuming a 10 \% Higgs boson decay branching fraction into \gammad. As a reference, the boundaries defining regions A, B, C and D are indicated in the figure by solid red lines. 

In order to take into account the small signal contamination in regions B, C and D, a likelihood-based ABCD method is used  for the background estimation in the SR.
It estimates the background in region A by performing a fit to the background and signal yields in the four regions. A likelihood function is formed from the product of four Poisson functions, one for each of the A, B, C, and D regions, describing signal and background expectations. The likelihood takes the form:
\begin{eqnarray}
& & \mathcal{L}(n_{\textrm{A}},n_{\textrm{B}},n_{\textrm{C}},n_{\textrm{D}} | s, b, \tau_{\textrm{B}}, \tau_{\textrm{C}}) = \prod_{i={\textrm{A,B,C,D}}} \frac{{\textrm{e}}^{-N_i} N_i^{n_i}}{n_i !} \quad , \nonumber
\end{eqnarray}
where $n_{\textrm{A}}$, $n_{\textrm{B}}$, $n_{\textrm{C}}$ and $n_{\textrm{D}}$ are the four observables that denote the number of events observed in each region in data. The $N_i$ are linear combinations of the signal and background expectation in each region, defined as follows:
\begin{eqnarray}
 & & N_{\textrm{A}} =  s + b \nonumber \\
 & & N_{\textrm{B}} = s \, \varepsilon_{\textrm{B}} + b \, \tau_{\textrm{B}}  \nonumber \\
 & & N_{\textrm{C}} =  s \, \varepsilon_{\textrm{C}} + b \, \tau_{\textrm{C}}  \nonumber \\
 & & N_{\textrm{D}} = s \, \varepsilon_{\textrm{D}} + b \, \tau_{\textrm{C}} \, /\ \tau_{\textrm{B}}   \nonumber 
\end{eqnarray}
where $s$ ($b$) is the signal (background) yield in region A, $\varepsilon_i$ is the signal contamination derived from MC simulation,  and $\tau_{\textrm{B}}$ and $\tau_{\textrm{C}}$ are the nuisance parameters that describe the ratio of the background expectation in the control region to the background expectation in the signal region. The $s$, $b$, $\tau_{\textrm{B}}$ and $\tau_{\textrm{C}}$ values are allowed to float in the fit to the four data regions. 

%% file: Systematics_v6.tex
\section{Systematic uncertainties}
\label{sec:syst}
The uncertainty in the ABCD-method background estimate is evaluated from the impact of a possible correlation between the ABCD variables in the SR. The correlation is evaluated using multi-jet MC events and validated in VR data. This effect is found to lead to a potential variation of less than 4\% in the background estimate. The size of this uncertainty is therefore considered negligible when compared to the statistical one and it is not included in the fit.

 The following effects are considered as possible sources of systematic uncertainty in the signal.
%
%
\paragraph{Luminosity}\mbox{} \\
The uncertainty in the combined 2015--2016 integrated luminosity is 2.1\%~\cite{ATLAS-CONF-2019-021}, obtained using the LUCID-2 detector~\cite{LUCID2} for the primary luminosity measurements.

%
%
\paragraph{Trigger}\mbox{} \\ 
The systematic uncertainty in the narrow-scan trigger efficiency is evaluated using a tag-and-probe method applied to $J/\psi \to \mu\mu$ events in data and simulation. 
The difference between the trigger efficiency in data and that in simulation is evaluated as a function of the opening angle between the two muons. The difference in the region \mydelr~<~0.05, corresponding to the \mydelr\ expected for signal, is taken as the uncertainty and is 6$\%$.
The systematic uncertainty in the tri-muon MS-only trigger efficiency is 5.8$\%$, taken from the analysis of 2012 data~\cite{EXOT-2013-22} since the algorithm has not undergone a major change since then. 
The systematic uncertainty in the CalRatio trigger efficiency is taken from Ref.~\cite{EXOT-2017-25} and is 2$\%$.
%
\paragraph{BDT shape} \mbox{} \\
The systematic uncertainty in the MC modelling of the input variables used for the BDT training is evaluated for both the \mLJ and the \hLJ. For the \mLJ, the data-to-MC ratio is computed for muon timing and $z_0$ BDT input variables using samples of $Z\rightarrow\mu\mu$ events. This comparison relies on the fact that, due to the high boost of low-mass dark photons and the high \mypt of signal muons, the muon $z_0$ and timing distributions are similar to those of prompt muons from $Z\rightarrow\mu\mu$. Muons from $Z$ boson decay are reconstructed using information only from the MS. The BDT is retrained using MC signal variables scaled to the data using these ratios, and the fit procedure is repeated. The resulting change in the final signal yield is taken as the systematic uncertainty, and its value is 3\%. The same procedure is used for the \hLJ, where the ratios of data to simulated distributions are computed from data and MC samples of multi-jet events. The resulting uncertainty is 14\%.
%
\paragraph{Muon reconstruction} \mbox{} \\
The systematic uncertainty in the single-\gammad reconstruction efficiency is evaluated using a tag-and-probe method applied to $J/\psi \to \mu\mu$ events in 2015 data and simulation. $J/\psi\rightarrow \mu\mu$ decays are selected, and the efficiency is evaluated as a function of the opening angle \mydelr between the two muons, for both the data and simulated $J/\psi$ decays. For low \mydelr values, the efficiency decreases due to the difficulty of reconstructing two tracks with small angular separation in the MS. The difference in $J/\phi \to \mu\mu$ reconstruction efficiency between simulation and data in the \mydelr~interval between 0 and 0.06 (where the \LJ samples are concentrated) amounts to 15$\%$, and is taken as the uncertainty.
%
\paragraph{Jet energy scale and jet energy resolution}\mbox{} \\
The jet energy scale and jet energy resolution introduce uncertainties in the signal yield of 1\% to 8\% and 1\% to 5\% respectively, depending on the signal process, where the processes with two dark photons are less affected. These uncertainties are calculated using the procedure detailed in Ref. \cite{PERF-2016-04}. Since the jets used in this analysis are required to have a low fraction of energy in the electromagnetic calorimeter, additional jet energy uncertainties are derived as a function of electromagnetic energy fraction as well as of pseudorapidity. These additional jet energy uncertainties are found to have an effect of up to 4\% on the signal yield, and are taken in quadrature with the regular jet energy uncertainties.
%
\paragraph{Effect of pile-up on \SumpT} \mbox{} \\
The presence of multiple collisions per bunch crossing affects the efficiency of the ID track isolation criterion  quantified in terms of \SumpT. The systematic uncertainty is evaluated by comparing \SumpT for muons from a sample of reconstructed $Z \to \mu\mu$ events in data with that in simulation, as a function of the number of interaction vertices in the event. The systematic uncertainty is evaluated as the maximum difference at the value of the selection requirement on \maxsumpt. It is found to be 5.1$\%$.
%

%


%% file: Results_v7.tex
\section{Results and interpretation}
\label{sec:results}

The observed numbers of events in the ABCD regions and the expected number of background events in the signal region A are summarised in Table~\ref{tab:abcd}. The expected number of background events in region A is estimated using the likelihood-based ABCD method, assuming no signal and not including the observed data in region A. The background estimate includes both the multi-jet and cosmic-ray background, where the former is obtained as described in Section~\ref{sec:bkg}, and the latter is estimated from the cosmic dataset. 
Both sources are included in the expected background given in Table~\ref{tab:abcd}.
The observed number of events in region A is in agreement with the predicted number of background events.

\begin{table}[h]
\centering
\caption{Observed numbers of events in the ABCD regions and expected number of background events in region A. In the estimate, the data in region A are not considered and the signal strength is fixed to zero.
The expected contribution from cosmic rays is included in all regions.}
\small
\begin{tabular}{|c|r|r|r|r@{\;$\pm$\;}l|r|}
\hline
	\LJ pair type    & B   & C   & D   &  \multicolumn{2}{c|}{ Expected A}    &  A   \\
\hline
\hline
	\mLJ--\mLJ  &  24  &   92  &   463 &  128  &26 (stat.) & 113  \\
\hline
	\mLJ--\hLJ & 8     &   2   &   45  & 177  &86 (stat.) &   179 \\
\hline
	\hLJ--\hLJ & 13    &   2   &   15  & 97  &48 (stat.) & 69 \\
\hline
\end{tabular}
\label{tab:abcd}
\end{table}

Table~\ref{tab:FRVZabcd} shows the expected number of signal events in region A for the FRVZ model parameters of Table~\ref{tab:ggh_FRVZ_param} and the following assumptions: a value of \BR$(H\rightarrow \fdd \fddbar)$~=~10\% for a Higgs boson with \mH~=~125~\GeV, which is not excluded by the current measurements~\cite{Aaboud:2019rtt}; 
a value of \BR$(H\rightarrow \fdd \fddbar)$~=~100\% and a production cross section of $\sigma~=~5$~pb for a BSM scalar boson with \mH~=~800~\GeV.

\begin{table}[h]
\centering
\caption{Expected numbers of signal events in region A.  A branching fraction value of \BR$(H\rightarrow \fdd \fddbar)$~=~10\% is assumed for \LJ production in the decay of the \mH~=~125~\mygev Higgs boson. For \LJ production in the decay of a \mH~=~800~\mygev BSM scalar boson, a value of \BR$(H\rightarrow \fdd \fddbar)$~=~100\% and a production cross section of $\sigma~=~5$~pb are assumed. Only statistical uncertainties are reported. 
}
\small
\begin{tabular}{| c| r@{\;$\pm$\;}l | r@{\;$\pm$\;}l | r@{\;$\pm$\;}l| r@{\;$\pm$\;}l|}
\hline
	\LJ pair type & \multicolumn{2}{c|}{\mH = 125 \mygev} &  \multicolumn{2}{c|}{\mH = 125 \mygev} &  \multicolumn{2}{c|}{ \mH = 800 \mygev}&  \multicolumn{2}{c|}{ \mH = 800 \mygev}\\
	& \multicolumn{2}{c|}{ \gghHtwogd}& \multicolumn{2}{c|}{\gghHfourgd}& \multicolumn{2}{c|}{\gghHtwogd} & \multicolumn{2}{c|}{\gghHfourgd}\\
\hline
\hline
	\mLJ--\mLJ &\phantom{ia} 639 & 25 & \phantom{ia} 519  & 23 &\phantom{ia} 610 & 87 &\phantom{ia}  660 & 91\\
\hline
	\mLJ--\hLJ & 74 & 9& 22  &5 & 1544 & 139  & 996 & 111 \\
\hline
	\hLJ--\hLJ & 8 & 3 &  \multicolumn{2}{c|}{0} & 560 & 84  &  336 & 65 \\
\hline
\end{tabular}
\label{tab:FRVZabcd}
\end{table}

\begin{figure}[h]
 \centering
 \includegraphics[width=.45\textwidth]{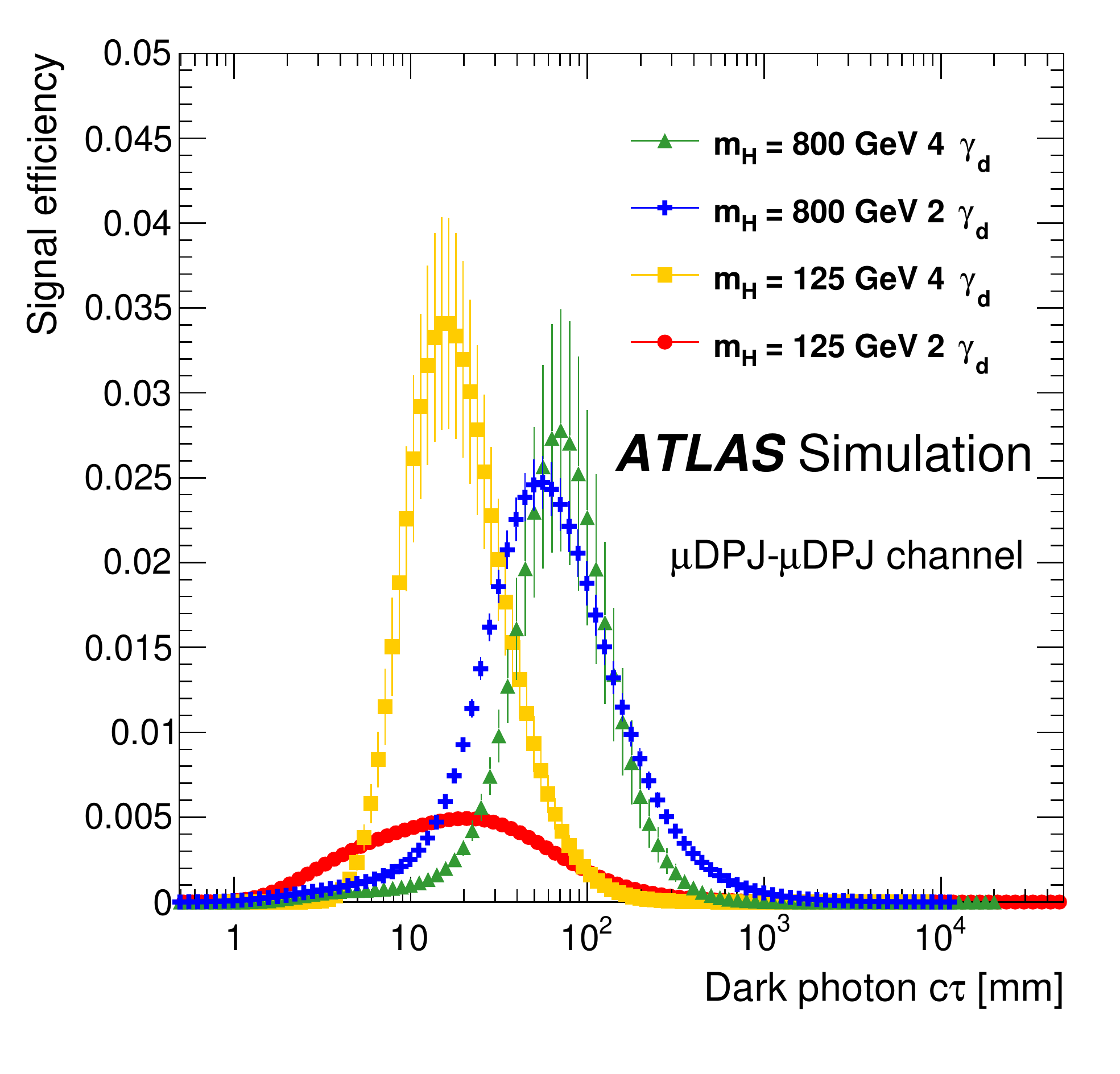}
 \includegraphics[width=.45\textwidth]{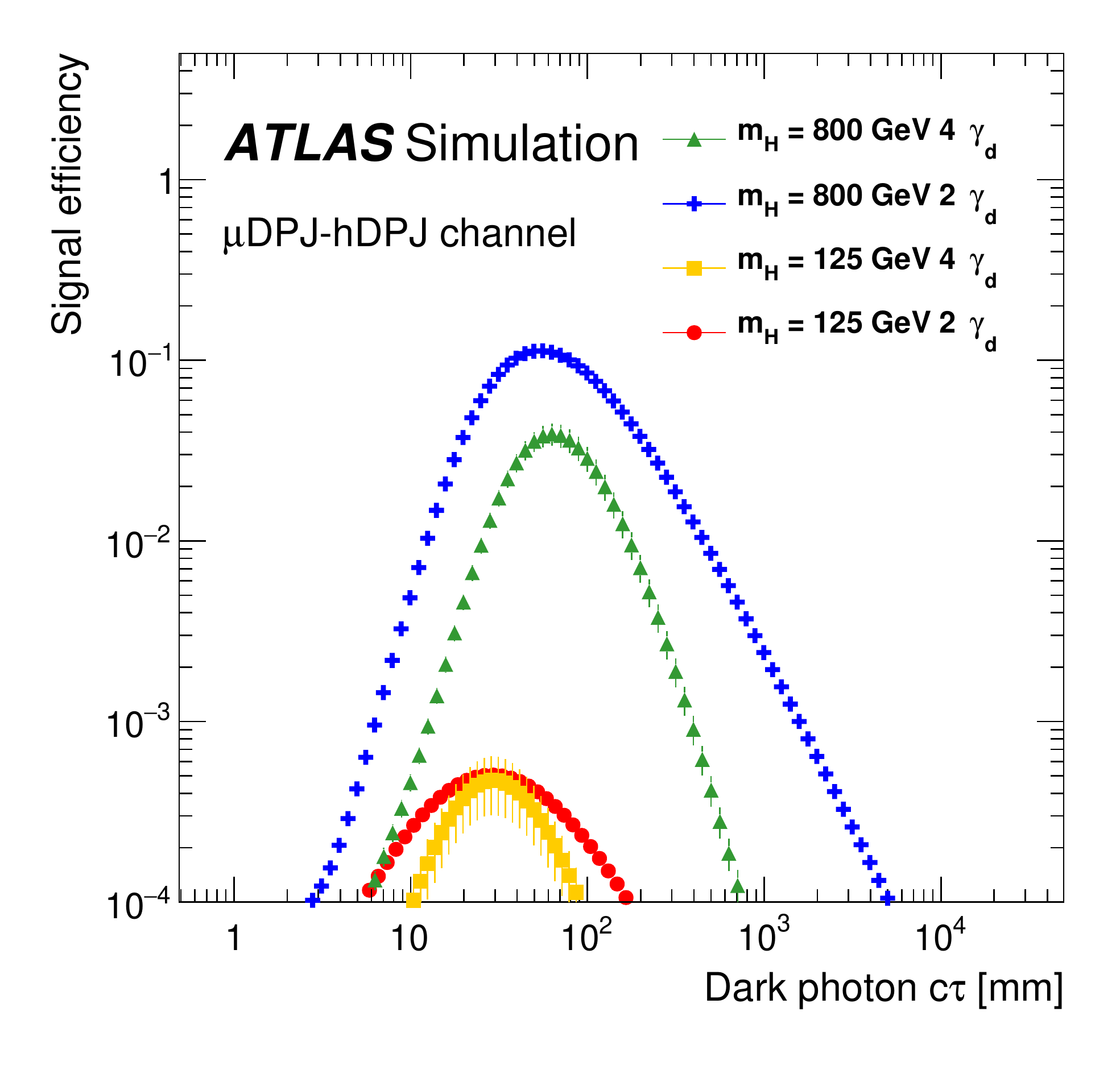}
 \includegraphics[width=.45\textwidth]{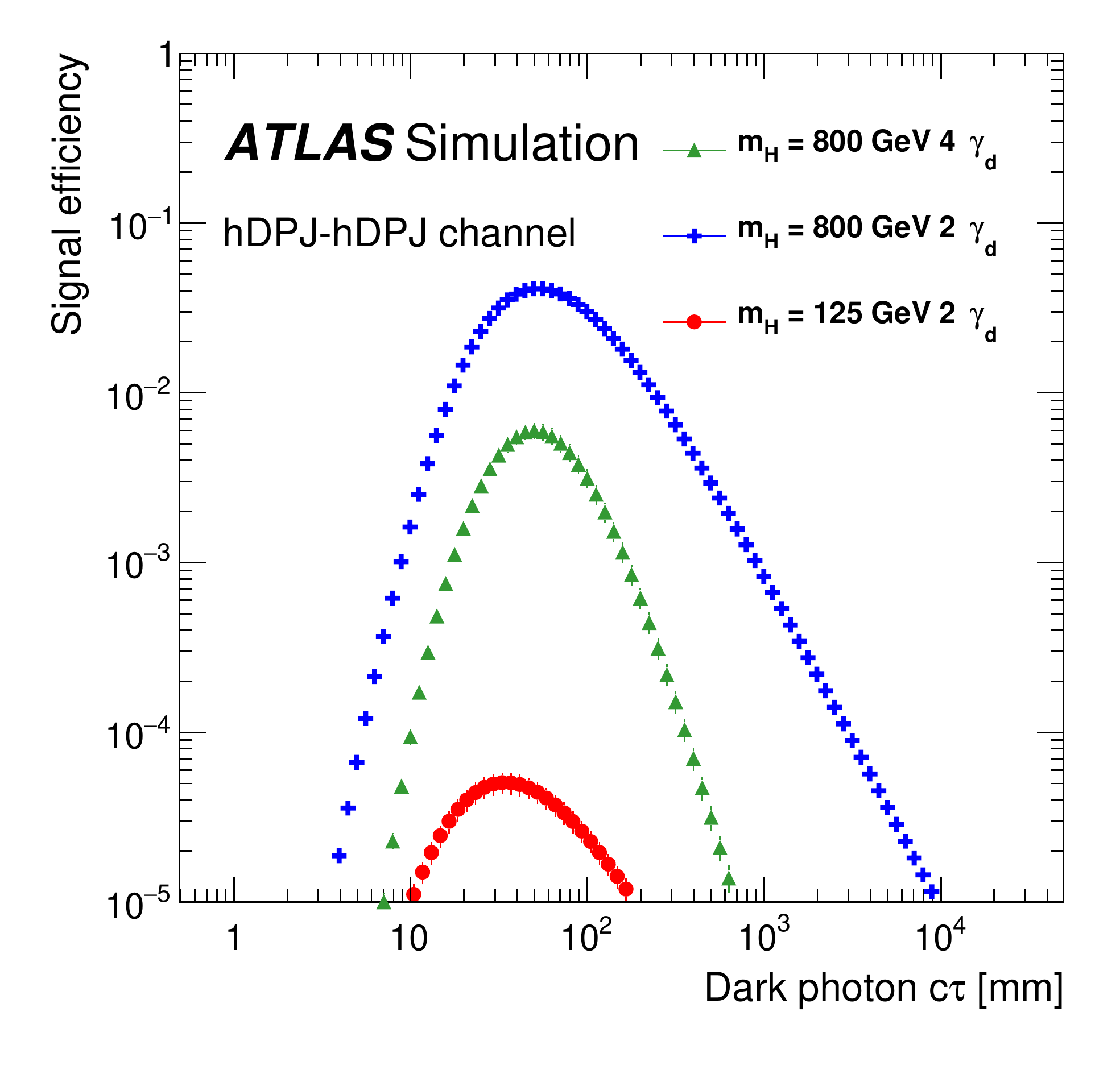}
 \caption{Extrapolated signal efficiencies as a function of proper decay length of the \gammad for the \gghHtwogd and \gghHfourgd processes  and for the three different channels: \mLJ--\mLJ (left), \mLJ--\hLJ (right) and \hLJ--\hLJ (bottom). 
 The signal efficiency in the \hLJ--\hLJ~channel for \mH~$=125$~\mygev \gghHfourgd process is small compared with the other channels and is not shown.
 The vertical bars represent the statistical uncertainties.}
 \label{fig:rescale}
 \end{figure}

 Upper limits on the production cross section times branching fraction (\sigmabr) as a function of the \gammad proper decay length are derived for the FRVZ \gghHtwogd and \gghHfourgd processes using the CL$_\text{s}$ method \cite{CLspaper}.
Since each signal sample was generated for a particular proper decay length, it is necessary to extrapolate the signal efficiency to other decay lengths to obtain limits as a function of \ctau.
This is achieved by applying to the $i$-th dark photon in the event a weight

 \begin{equation*}
  w_i(t_i) = \frac{\tau_\textrm{ref}}{\textrm{e}^{-t_i/\tau_\textrm{ref}}} \cdot \frac{\textrm{e}^{-t_i/\tau_\textrm{new}}}{\tau_\textrm{new}},
\end{equation*}

 where $\tau_\textrm{ref}$ is the lifetime with which the event sample was simulated, $\tau_\textrm{new}$ is the lifetime for which it is weighted, and $t_i$ is the proper decay time of the $i$-th dark photon. Each event is weighted by the product of the individual dark-photon weights. The weighted sample is used to evaluate the signal efficiency for $\tau_{\textrm{new}}$. 
Figure~\ref{fig:rescale} shows the extrapolated signal efficiency for the \gghHtwogd and \gghHfourgd processes as a function of \ctau of the dark photon in the \mLJ--\mLJ, \mLJ--\hLJ and \hLJ--\hLJ channels. The tri-muon MS-only trigger has a lower efficiency for the \gghHtwogd process with a \mH~=~125~\mygev Higgs boson than for the other processes, resulting in a lower signal efficiency in the \mLJ--\mLJ channel. 
The \mypt requirements of the CalRatio trigger are not optimal for selecting jets produced by \gammad decays in the processes with a \mH~=~125~\mygev Higgs boson, resulting in a signal efficiency below 1\% in the \hLJ--\hLJ channel. The muon narrow-scan trigger helps to recover some efficiency in the \mLJ--\hLJ channel for these processes.

\begin{figure}[h]
 \centering
 \includegraphics[width=.49\textwidth]{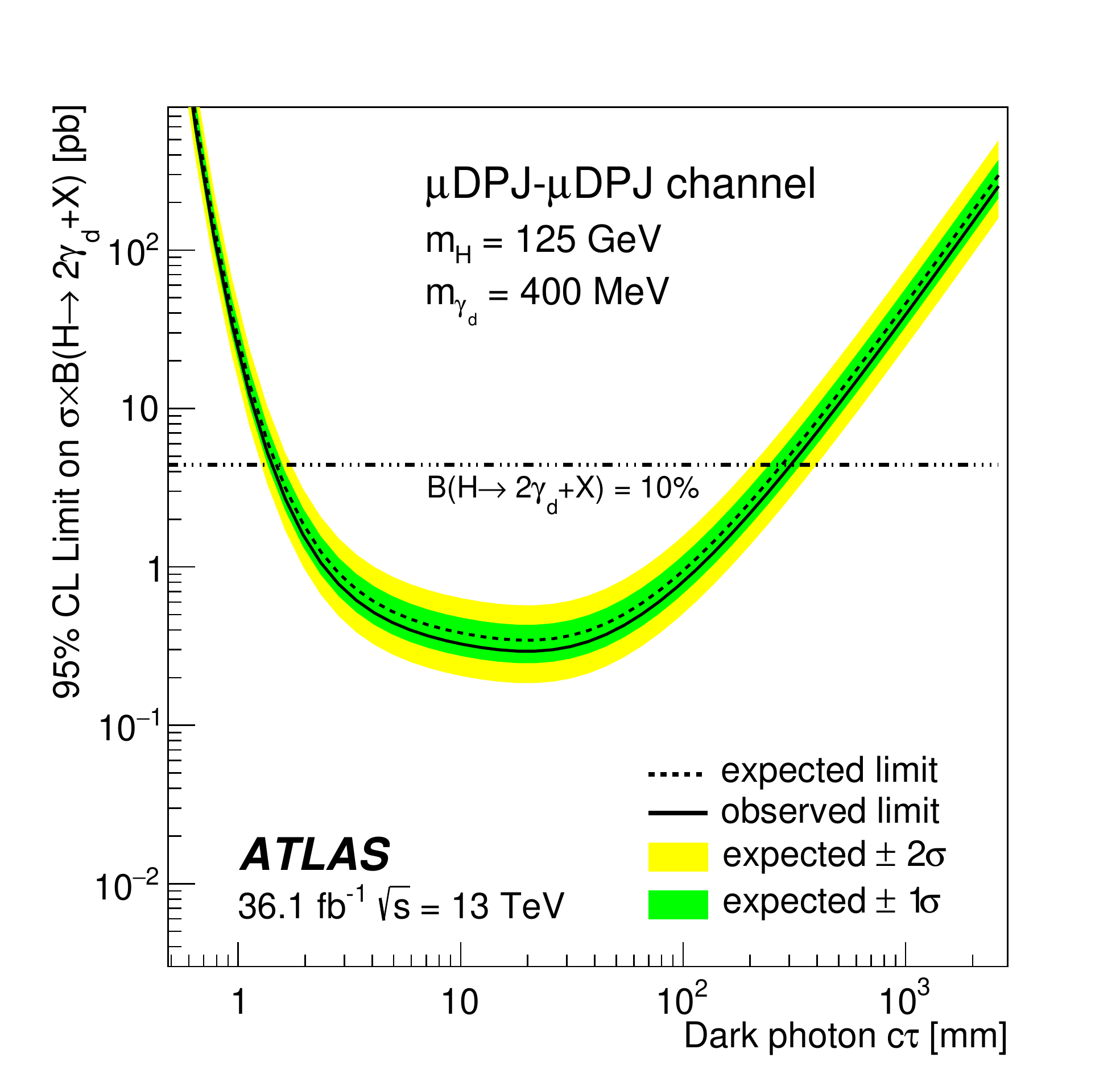}
 \includegraphics[width=.49\textwidth]{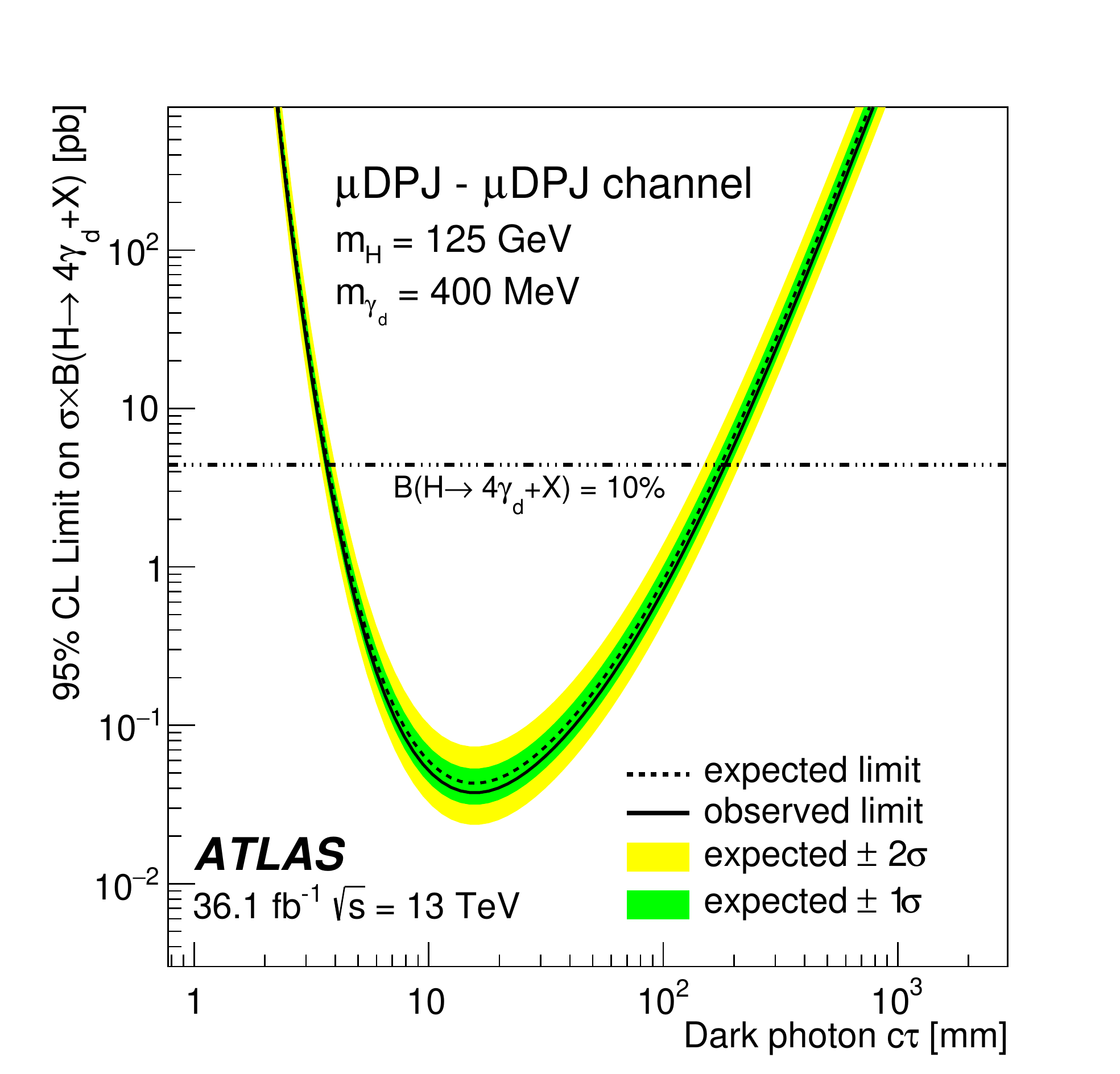}
 \caption{Upper limits at 95$\%$ CL on the cross section times branching fraction for the processes \gghHtwogd (left) and \gghHfourgd (right) in the \mLJ--\mLJ final states for \mH~$=125$~\mygev. The horizontal lines correspond to the cross section times branching fraction for a value of the branching fraction of the Higgs boson decay into dark fermions of 10\%.}
 \label{fig:limits00}
 \end{figure}

\begin{figure}[h]
 \centering
 \includegraphics[width=.49\textwidth]{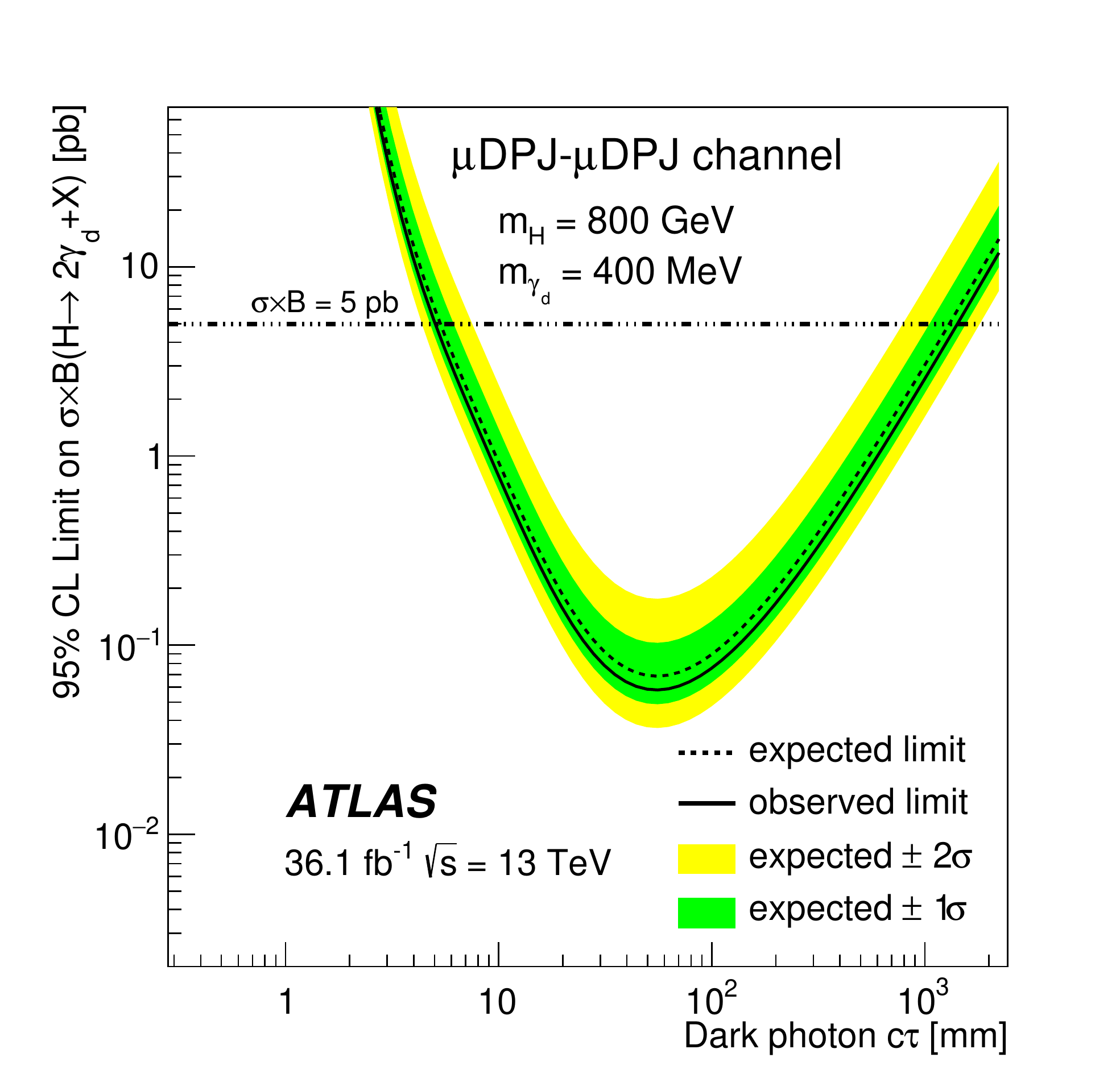}
 \includegraphics[width=.49\textwidth]{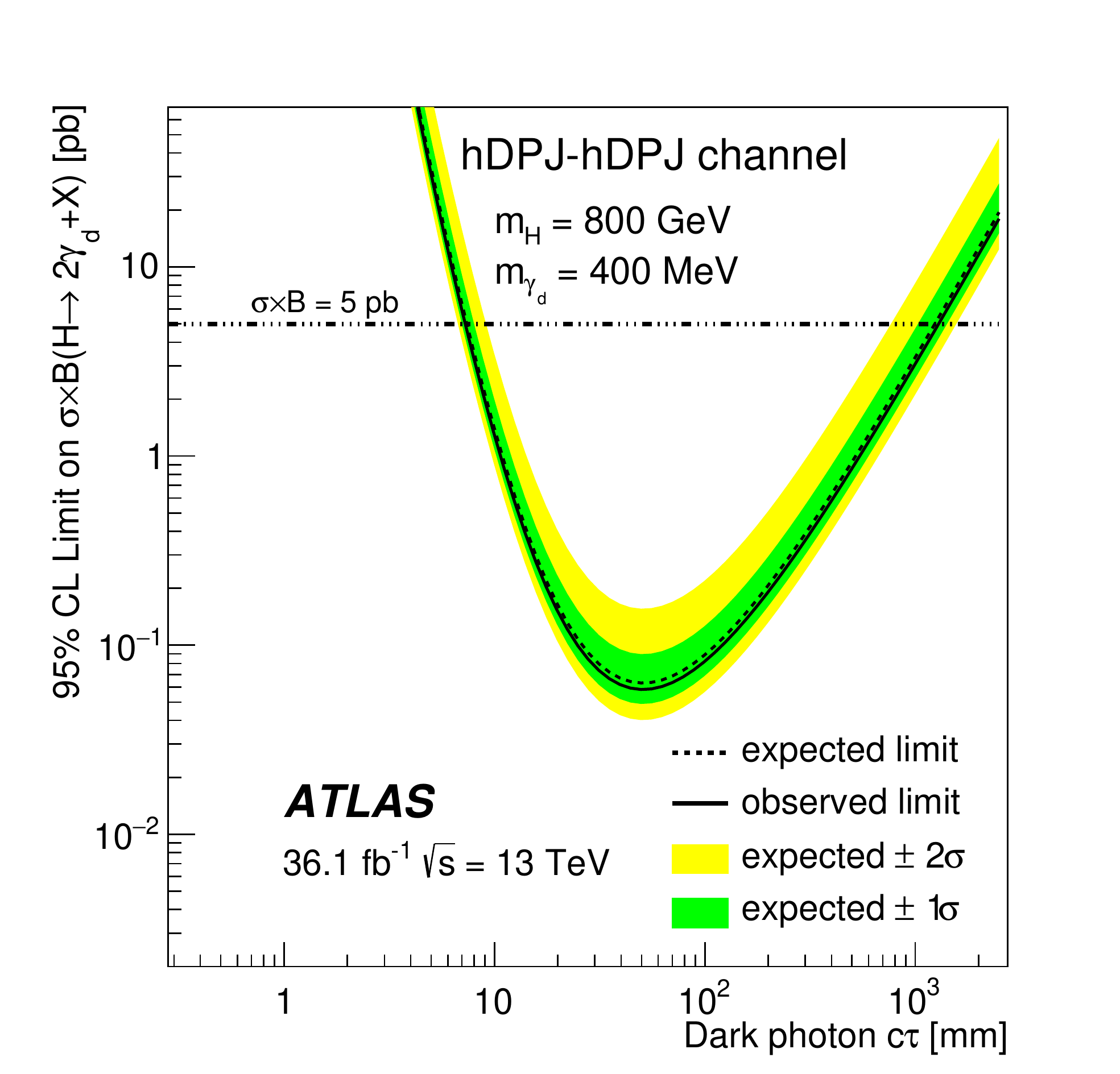}
 \caption{Upper limits at 95$\%$ CL on the cross section times branching fraction for the process \gghHtwogd, where $H$ is an 800~\mygev BSM Higgs boson, in the \mLJ--\mLJ (left) and \hLJ--\hLJ (right) final states. 
 The horizontal lines correspond to a cross section times branching fraction of 5~pb.}
 \label{fig:limits800}
 \end{figure}

\begin{figure}[ht!]
\centering
\includegraphics[width=120mm]{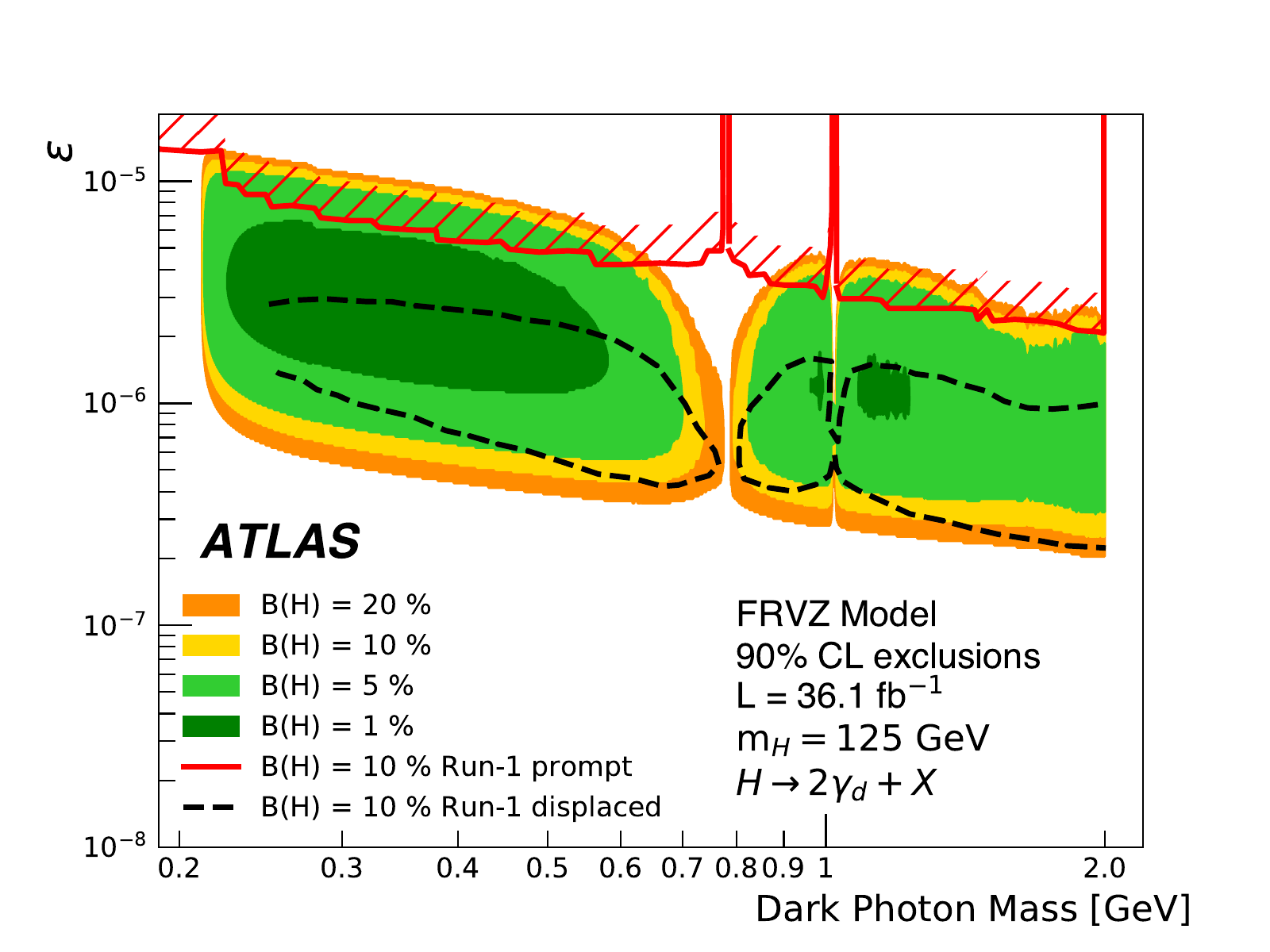}
\caption{ The 90$\%$ CL exclusion regions for the decay \gghHtwogd of the Higgs boson as a function of the \gammad mass and of the kinetic mixing parameter $\epsilon$. 
These limits are obtained assuming the FRVZ model with decay branching fractions of the Higgs boson into \gammad between 1\% and 20\%, and the NNLO Higgs boson production cross sections via gluon--gluon fusion.
	The figure also shows excluded regions with decay branching fraction of the Higgs boson into \gammad of 10\% from the run-1 ATLAS displaced ~\cite{EXOT-2013-22} (black line) and prompt~\cite{EXOT-2014-09} (red line) dark-photon jets searches.
}
\label{fig:DMexclusion}
\end{figure}

The observed 95$\%$ CL cross-section upper limits in the \mLJ--\mLJ~ channel for the \gghHtwogd~ and \gghHfourgd~ processes are presented in Figure~\ref{fig:limits00} for \mH~=~125~\mygev. The 95$\%$ CL exclusion limits in the \mLJ--\mLJ~ and \hLJ--\hLJ~ channels for the process \gghHtwogd~ are presented in Figure~\ref{fig:limits800} for \mH~=~800~\mygev.
The figures also show the expected limits obtained from the likelihood-based ABCD method, using the background estimate derived from the background-only fit using data in the four regions.
Excluded \ctau ranges are summarised in Table~\ref{tab:limits}, assuming \BR$(H\rightarrow \fdd \bar{\fdd})~=~10\%$ for the Higgs boson with \mH~=~125~\mygev and \BR$(H\rightarrow \fdd \bar{\fdd})~=~100\%$ for the BSM Higgs boson, with subsequent decay of the $\fdd$ and $\bar{\fdd}$ giving rise to the production of two or four dark photons.
\renewcommand{\arraystretch}{1.2}
\begin{table}[t!]
\caption{Ranges of \gammad \ctau excluded at 95$\%$ CL for \gghHtwogd and \gghHfourgd. A branching fraction value of \BR$(H\rightarrow \fdd \bar{\fdd})$~=~10\% is assumed for \LJ production in the decay of a \mH~=~125~\GeV\ Higgs boson. For \LJ production in the decay of a \mH~=~800~\mygev BSM scalar boson, a value of \BR$(H\rightarrow \fdd \bar{\fdd})$~=~100\% and a production cross section of $\sigma~=~5$~pb are assumed.
}
\small
\centering
\begin{tabular}{| c | c | c | c| c|}
\hline
 & Excluded \ctau $[\textrm{mm} ]$ & Excluded \ctau $[\textrm{mm} ]$ & Excluded \ctau $[\textrm{mm} ]$ & Excluded \ctau $[\textrm{mm} ]$\\
Model  & \mH = 125 \mygev & \mH = 125 \mygev &  \mH = 800 \mygev&  \mH = 800 \mygev\\
& \gghHtwogd & \gghHfourgd & \gghHtwogd & \gghHfourgd\\
\hline
\hline
\mLJ--\mLJ  & 1.5 $\leq$ \ctau $\leq$ 307 & 3.7 $\leq$ \ctau $\leq$ 178 &  5.0 $\leq$ \ctau $\leq$ 1420 & 10.5  $\leq$ \ctau $\leq$ 312 \\
\hline
\mLJ--\hLJ  &   --  &   -- & 7.2 $\leq$ \ctau $\leq$ 1234  & 14.5 $\leq$ \ctau $\leq$ 334 \\
\hline
\hLJ--\hLJ  &   --  &   -- & 7.3 $\leq$ \ctau $\leq$1298 &  13.6 $\leq$ \ctau $\leq$ 231 \\
\hline
%
\end{tabular}
\label{tab:limits}
\end{table}
The results for the \mLJ--\mLJ~ channel is also interpreted in terms of the kinetic mixing parameter $\epsilon$ and \gammad mass, shown in Figure~\ref{fig:DMexclusion} as exclusion contours.
These limits assume four possible values of the Higgs boson decay branching fractions into \gammad, ranging from 1\%~to~20\%, and the NNLO gluon--gluon fusion Higgs boson production cross section. The \gammad detection efficiency for a \gammad mass of $0.4$~\mygev is used for the mass interval 0.25--2~\mygev, as the detection efficiency is constant  throughout this interval~\cite{EXOT-2013-22}.
The decay branching fraction variations as a function of the \gammad mass are estimated and included in the 90$\%$ CL exclusion region evaluations~\cite{Batell:2009yf}. 
The low sensitivity in the \hLJ--\hLJ channel prevents the exclusion of the mass regions where the \gammad decays into hadronic resonances: \gammad mass regions around 0.8 and 1.0~\mygev, where the \gammad decays into the $\rho$, $\omega$, and $\phi$ resonances. 
Figure~\ref{fig:DMexclusion} also shows previous exclusions for a Higgs boson decay branching fractions into \gammad of 10\%  from a search for displaced dark-photon jets~\cite{EXOT-2013-22} and prompt dark-photon jets~\cite{EXOT-2014-09} at ATLAS. 
The search of Ref~\cite{EXOT-2013-22}, which explored the same region probed by this analysis, is slightly more sensitive in the region of high \gammad mass and low $\epsilon$. 
This is due to inclusion of dark-photon jets with both muon and hadron constituents, which are not used in the current analysis. 
The search of Ref.~\cite{EXOT-2014-09} excluded high $\epsilon$ values (shorter lifetimes), a region complementary to this analysis.

%% file: Conclusions_v4.tex
\FloatBarrier
\section{Conclusions}
\label{sec:conclusions}
The ATLAS detector at the LHC is used to search for the production of displaced dark-photon jets in a sample of $pp$ collisions at $\sqrt{s} =13$~\mytev corresponding to an integrated luminosity of \intlumi. 
No significant excess of events compared with the background expectation is observed, and 95\% confidence-level upper limits are set on the production cross section times branching fraction of scalar bosons that decay into dark photons according to the FRVZ model. The upper limits are computed as a function of the proper decay length \ctau~of the dark photon \gammad.
In addition to the increase in integrated luminosity and centre-of-mass energy, improvements in background suppression and the exploitation of hadronic \gammad decays result in increased sensitivity compared with the ATLAS search using 8~\TeV\ $pp$ data.
In the pure muonic channel, assuming a branching ratio \BR$(H\rightarrow~2(4)\gamma_\textrm{d}+X)$~=~10\% for $m_H~=~125$~\mygev, decays of dark photons are excluded at 95\% CL for $c\tau~\in~[1.5, 307$]~mm and $c \tau~\in~[3.7, 178]$~mm for production of two and four dark photons, respectively.  For $m_H~=~800$~\mygev, assuming  $\sigma~\times$~\BR$(H\rightarrow~2(4)\gamma_\textrm{d}+X)$~=~5~pb, the excluded regions are $c \tau~\in~[5, 1420]$~mm and $c\tau~\in~[10.5, 312]$~mm. In the pure hadronic channel, the $m_H~=~800$~\mygev excluded regions become $c\tau~\in~[7.3, 1298]$~mm and $c\tau~\in~[13.6, 231]$~mm.

The results for \gghHtwogd, when $H$ is the Higgs boson, are also interpreted as 90\% confidence-level limits on the kinetic mixing parameter as a function of the  dark-photon  mass. These results improve upon the constraints set in previous LHC searches.

%% file: Acknowledgements.tex

We thank CERN for the very successful operation of the LHC, as well as the
support staff from our institutions without whom ATLAS could not be
operated efficiently.

We acknowledge the support of ANPCyT, Argentina; YerPhI, Armenia; ARC, Australia; BMWFW and FWF, Austria; ANAS, Azerbaijan; SSTC, Belarus; CNPq and FAPESP, Brazil; NSERC, NRC and CFI, Canada; CERN; CONICYT, Chile; CAS, MOST and NSFC, China; COLCIENCIAS, Colombia; MSMT CR, MPO CR and VSC CR, Czech Republic; DNRF and DNSRC, Denmark; IN2P3-CNRS, CEA-DRF/IRFU, France; SRNSFG, Georgia; BMBF, HGF, and MPG, Germany; GSRT, Greece; RGC, Hong Kong SAR, China; ISF and Benoziyo Center, Israel; INFN, Italy; MEXT and JSPS, Japan; CNRST, Morocco; NWO, Netherlands; RCN, Norway; MNiSW and NCN, Poland; FCT, Portugal; MNE/IFA, Romania; MES of Russia and NRC KI, Russian Federation; JINR; MESTD, Serbia; MSSR, Slovakia; ARRS and MIZ\v{S}, Slovenia; DST/NRF, South Africa; MINECO, Spain; SRC and Wallenberg Foundation, Sweden; SERI, SNSF and Cantons of Bern and Geneva, Switzerland; MOST, Taiwan; TAEK, Turkey; STFC, United Kingdom; DOE and NSF, United States of America. In addition, individual groups and members have received support from BCKDF, CANARIE, CRC and Compute Canada, Canada; COST, ERC, ERDF, Horizon 2020, and Marie Sk{\l}odowska-Curie Actions, European Union; Investissements d' Avenir Labex and Idex, ANR, France; DFG and AvH Foundation, Germany; Herakleitos, Thales and Aristeia programmes co-financed by EU-ESF and the Greek NSRF, Greece; BSF-NSF and GIF, Israel; CERCA Programme Generalitat de Catalunya, Spain; The Royal Society and Leverhulme Trust, United Kingdom. 

The crucial computing support from all WLCG partners is acknowledged gratefully, in particular from CERN, the ATLAS Tier-1 facilities at TRIUMF (Canada), NDGF (Denmark, Norway, Sweden), CC-IN2P3 (France), KIT/GridKA (Germany), INFN-CNAF (Italy), NL-T1 (Netherlands), PIC (Spain), ASGC (Taiwan), RAL (UK) and BNL (USA), the Tier-2 facilities worldwide and large non-WLCG resource providers. Major contributors of computing resources are listed in Ref.~\cite{ATL-GEN-PUB-2016-002}.

%% file: atlas_authlist.tex
 
\begin{flushleft}
{\Large The ATLAS Collaboration}

\bigskip

G.~Aad$^\textrm{\scriptsize 102}$,    
B.~Abbott$^\textrm{\scriptsize 129}$,    
D.C.~Abbott$^\textrm{\scriptsize 103}$,    
A.~Abed~Abud$^\textrm{\scriptsize 71a,71b}$,    
K.~Abeling$^\textrm{\scriptsize 53}$,    
D.K.~Abhayasinghe$^\textrm{\scriptsize 94}$,    
S.H.~Abidi$^\textrm{\scriptsize 167}$,    
O.S.~AbouZeid$^\textrm{\scriptsize 40}$,    
N.L.~Abraham$^\textrm{\scriptsize 156}$,    
H.~Abramowicz$^\textrm{\scriptsize 161}$
% ...
\bigskip
\\

% ...
$^{40}$Niels Bohr Institute, University of Copenhagen, Copenhagen; Denmark.\\
% ...
$^{53}$II. Physikalisches Institut, Georg-August-Universit\"{a}t G\"ottingen, G\"ottingen; Germany.\\
% ...
$^{71}$$^{(a)}$INFN Sezione di Pavia;$^{(b)}$Dipartimento di Fisica, Universit\`a di Pavia, Pavia; Italy.\\
% ...
$^{94}$Department of Physics, Royal Holloway University of London, Egham; United Kingdom.\\
% ...
$^{102}$CPPM, Aix-Marseille Universit\'e, CNRS/IN2P3, Marseille; France.\\
$^{103}$Department of Physics, University of Massachusetts, Amherst MA; United States of America.\\
% ...
$^{129}$Homer L. Dodge Department of Physics and Astronomy, University of Oklahoma, Norman OK; United States of America.\\
% ...
$^{156}$Department of Physics and Astronomy, University of Sussex, Brighton; United Kingdom.\\
% ...
$^{161}$Raymond and Beverly Sackler School of Physics and Astronomy, Tel Aviv University, Tel Aviv; Israel.\\
% ...
$^{167}$Department of Physics, University of Toronto, Toronto ON; Canada.\\
% ...
$^{a}$ Also at Borough of Manhattan Community College, City University of New York, New York NY; United States of America.\\
$^{b}$ Also at CERN, Geneva; Switzerland.\\
% ...

\end{flushleft}
